\documentclass{svjour2}
\usepackage{amscd}
\usepackage{amsfonts}
\usepackage{amssymb}
\usepackage{amsmath}
\usepackage{dcolumn}
\usepackage[T1]{fontenc}
\usepackage[latin1]{inputenc}
\usepackage{graphicx}
\usepackage{graphics}
\usepackage{color}
\usepackage{latexsym}
\usepackage{amsfonts}
\usepackage{undertilde}
\usepackage{color}
\usepackage{hyperref}
\usepackage{wrapfig}
\usepackage{setspace}
\def\>{\rangle}
\def\<{\langle}
\def\n{\nonumber}

\def\sc{\scriptsize}

\begin{document}

\title{Non-equilibrium dynamics in the quantum Brownian oscillator and the second law of thermodynamics}

\author{Ilki Kim}

\institute{I. Kim \at
              Department of Physics, North Carolina A$\&$T State University, Greensboro, North Carolina 27411, USA\\
              Tel.: -1-336-285-2122\\
              Fax: -1-336-256-0815\\
              \email{hannibal.ikim@gmail.com}}

\date{\today}

\maketitle

\begin{abstract}
We initially prepare a quantum linear oscillator weakly coupled to a
bath in equilibrium at an arbitrary temperature. We disturb this
system by varying a Hamiltonian parameter of the coupled oscillator,
namely, either its spring constant or mass according to an arbitrary
but pre-determined protocol in order to perform external work on it.
We then derive a closed expression for the reduced density operator
of the coupled oscillator along this non-equilibrium process as well
as the exact expression pertaining to the corresponding quasi-static
process. This immediately allows us to analytically discuss the
second law of thermodynamics for non-equilibrium processes. Then we
derive a Clausius {\em in}equality and obtain its validity
supporting the second law, as a consistent generalization of the
Clausius equality valid for the quasi-static counterpart, introduced
in \cite{KIM10}. 
\end{abstract}

%%%%%%%%%%%%%%%%%%%%%%%%%%%%%%%%%%%%%%%%%%%%%%%%%%%%%%%%%%%%%%%%%%%%%%%%%%%%%
\section{Introduction}
%%%%%%%%%%%%%%%%%%%%%%%%%%%%%%%%%%%%%%%%%%%%%%%%%%%%%%%%%%%%%%%%%%%%%%%%%%%%%
%
Over the past several decades there have been remarkable
breakthroughs in experimental techniques for probing non-classical
behaviors of small-scale quantum objects coupled to quantum
environments (see, e.g., \cite{CAP05}). Correspondingly, a more
sophisticated theoretical understanding of the thermodynamic nature
of such systems, especially in the low temperature regime where
quantum effects are dominant, has been substantially demanded. In
contrast to common quantum statistical mechanics, which is
intrinsically based on a vanishingly small coupling between system
and environment, the finite coupling strength between them in the
quantum regime causes some subtleties that must be recognized.

At the heart of the aforesaid ``quantum thermodynamics''
\cite{MAH04}, the second law of thermodynamics, assumed to be
inviolable by the scientific community for over a century
\cite{CAL85}, has been confronted by challenges with considerable
interest, and its absolute status has even come under increased
scrutiny \cite{CAP05,MAH04,SHE02}. In fact, this fundamental law of
nature has extensively been theoretically studied particularly in
the scheme of a quantum harmonic oscillator linearly coupled to an
independent-oscillator model of a heat bath (quantum Brownian
oscillator) in equilibrium at a (low) temperature $T$, mainly due to
its mathematical manageability.

A short overview of recent results either pro or contra the validity
of the quantum second law has been given in \cite{KIM10}; and the
final result therein was that a properly defined Clausius inequality
$\delta{\mathcal Q}_{\text{\sc eff}}^{\star} \leq T_{\text{\sc
eff}}^{\star}\,dS_N$ representing the second law is valid without
any previously argued violation in form of $\delta{\mathcal Q}
\not\leq T\,dS$ at $T \to 0$. In the above inequality, $S_N$
represents the von-Neumann entropy of the coupled oscillator, which
is, interestingly enough, identical to the thermal entropy of an
{\em uncoupled} effective oscillator in equilibrium. And
$\delta{\mathcal Q}_{\text{\sc eff}}^{\star}$ is a heat exchanged
between (weakly coupled) effective oscillator and bath, and
$T_{\text{\sc eff}}^{\star}$ a well-defined effective equilibrium
temperature. This effective temperature differs from the temperature
$T$ of the total system (oscillator plus bath) especially in the
strong-coupling limit, where the total-system temperature is in fact
not well-defined as an equilibrium temperature of the system
oscillator since the reduced equilibrium density operator of the
oscillator $\hat{H}_s$ is not any longer in form of the canonical
thermal state $\hat{\rho}_{\beta} \propto e^{-\beta \hat{H}_s}$ with
$\beta = 1/(k_B\,T)$ \cite{KIM10}. This discrepancy between these
two equilibrium temperatures is, of course, caused by the extra work
(or energy) needed to couple an (uncoupled) oscillator to a bath
\cite{COT11}. As a result, it may be legitimate to say that the
original form of the Clausius inequality for the coupled oscillator
in terms of the temperature $T$ is not well-defined rather than
being violated.

However, the entire discussion of the quantum second law has so far
been restricted to that for thermal equilibrium states, accordingly,
in form of the Clausius equality. On the other hand, there has
recently been an interesting result for a generalized Clausius
inequality for non-equilibrium quantum processes, but restricted to
isolated quantum systems (initially prepared at a thermal
equilibrium state) \cite{LUT10}. In this paper we extend the above
discussions into far-from-equilibrium processes in the scheme of
quantum Brownian oscillator as a prototype of open quantum systems,
but in the weak-coupling limit only since the exact treatment of the
non-equilibrium processes in the strong-coupling limit is pretty
much hopeless to leading to useful quantities in closed form to be
needed for our discussion [cf. Eqs.
(\ref{eq:reform_of_total_density_matrix1})-(\ref{eq:reduced_density_operator_general_treatment1})].
To do so, we consider the quantum oscillator with time-dependent
Hamiltonian parameters, which finally enables us to derive a
generalized Clausius {\em in}equality without any violation in
variation of the Hamiltonian parameters as our central result. In
fact, the time-dependent quantum oscillator has been studied by many
authors for last about 60 years, initiated by Husimi \cite{HUS53}.
In this paper we appeal to the method of quantum Liouville equation
in order to directly obtain the time-dependent density operator of
the system in consideration. This approach differs from that
developed by Husimi, which is based on the fact that the
Schr\"{o}dinger equation for an isolated linear oscillator, rather
than coupled to a bath, can reduce to a system of classical
equations for the oscillator.

The general layout of this paper is the following. In Sect.
\ref{sec:basics} we review the general results of quantum Brownian
oscillator needed for our later discussions. In Sect.
\ref{sec:time-dpt-density-op} we consider non-equilibrium processes
and then derive a closed expression for the time-dependent reduced
density operator of the oscillator weakly coupled to a bath at an
arbitrary time along the processes. In Sect.
\ref{sec:instantaneous-eigenvalue} the same discussion will take
place for the corresponding quasi-static processes. Next the second
law of thermodynamics for non-equilibrium processes will be
systematically discussed in Sect. \ref{sec:thermodynamics}. Finally
we give the concluding remarks of this paper in Sect.
\ref{sec:conclusion}.

%%%%%%%%%%%%%%%%%%%%%%%%%%%%%%%%%%%%%%%%%%%%%%%%%%%%%%%%%%%%%%%%%%%%%%%%%%%%%
\section{Basics of quantum Brownian oscillator}\label{sec:basics}
%%%%%%%%%%%%%%%%%%%%%%%%%%%%%%%%%%%%%%%%%%%%%%%%%%%%%%%%%%%%%%%%%%%%%%%%%%%%%
%
The quantum Brownian oscillator under investigation is described by
the model Hamiltonian (Caldeira-Leggett model) \cite{WEI08,ING98}
\begin{equation}\label{eq:total_hamiltonian1}
    \hat{H}_0\; =\; \hat{H}_s\, +\, \hat{H}_b\, +\, \hat{H}_{sb}\,,
\end{equation}
where a system linear oscillator, a bath, and a system-bath
interaction are
\begin{eqnarray}
    &\displaystyle \hat{H}_s\, =\, \frac{\hat{p}^2}{2 M} +
    \frac{k_0}{2}\,\hat{q}^2\;\;\; ;\;\;\;
    \hat{H}_b\, =\, \sum_{j=1}^N \left(\frac{\hat{p}_j^2}{2 m_j} +
    \frac{k_j}{2}\,\hat{x}_j^2\right)&\label{eq:total_hamiltonian2}\\
    &\displaystyle \hat{H}_{sb}\, =\, -\hat{q} \sum_{j=1}^N c_j\,\hat{x}_j\, +\, \hat{q}^2
    \sum_{j=1}^N \frac{c_j^2}{2\,k_j}\,,&\label{eq:total_hamiltonian21}
\end{eqnarray}
respectively. Here the coupling strengths $c_j$, and the spring
constants $k_0 = M \omega_0^2$ and $k_j = m_j \omega_j^2$. The total
system $\hat{H}_0$ is assumed to be within the canonical thermal
equilibrium state $\hat{\rho}_{\beta} = e^{-\beta
\hat{H}_0}/Z_{\beta}$, in form of a non-separable state
($\not\propto \hat{\rho}_s \otimes \hat{\rho}_b$) due to the
interaction $\hat{H}_{sb}$, where the partition function $Z_{\beta}
= \text{Tr}\,e^{-\beta \hat{H}_0}$. The second term of the
interaction $\hat{H}_{sb}$, proportional to $\hat{q}^2$, was
introduced in order to protect the pre-determined frequency
($\omega_0$) of the system oscillator $\hat{H}_s$ from its
modification induced by the system-bath coupling (the first term
linear in $\hat{x}_j$) \cite{WEI08}. Here the system and the bath
effectively share the energy in the coupling term, especially in the
strong-coupling limit ($c_j \gg 1$), and so it is in fact not
completely clear whether this energy should be interpreted as
belonging to the system or to the bath \cite{JAR04}. Therefore,
without the above second term, the internal energy $U_s =
\mbox{Tr}\,(\hat{H}_s\,\hat{\rho}_{\beta})$ of the coupled
oscillator alone, with its unique frequency $\omega_0$, as well as
its reduced density operator would not be well-defined [cf. Eqs.
(\ref{eq:density_matrix1}) and
(\ref{eq:hamiltonian_internal_energy1})]. In fact, from the
Heisenberg equations of motion for $\hat{q}$ and $\hat{p}$ we can
derive the quantum Langevin equation without the frequency shift as
\cite{WEI08,ING98}
\begin{equation}\label{eq:eq_of_motion1}
    M\,\ddot{\hat{q}}(t) \, +\,
    M \int_0^t d\tau\, \gamma(t -\tau)\, \dot{\hat{q}}(\tau)\,
    +\, M \omega_0^2\,\hat{q}(t)\; =\; \hat{\xi}(t)\,,
\end{equation}
where the damping kernel and the noise operator are, respectively,
given by
\begin{eqnarray}
    \gamma(t) &=& \frac{1}{M} \sum_{j=1}^N
    \frac{c_j^2}{m_j\,\omega_j^2}\,\cos(\omega_j\,t)\\
    \hat{\xi}(t) &=& -M \gamma(t)\,\hat{q}(0)\, +\,
    \sum_{j=1}^N c_j \left\{\hat{x}_j(0)\,\cos(\omega_j\,t)\, +\,
    \frac{\hat{p}_j(0)}{m_j\,\omega_j}\,\sin(\omega_j\,t)\right\}\,.\label{eq:damping_kernel1}
\end{eqnarray}
Here as required, $\<\hat{\xi}(t)\>_{\rho_{b'}} =
\text{Tr}\,\{\hat{\xi}(t)\,\hat{\rho}_{b'}\} = 0$, in which the
shifted bath state $\hat{\rho}_{b'} = e^{-\beta\,(\hat{H}_b +
\hat{H}_{sb})}/Z^{(b')}_{\beta}$ with the corresponding partition
function $Z^{(b')}_{\beta}$, and the noise correlation \cite{HAE05}
\begin{equation}\label{eq:qm_noise_correlation1}
    \frac{1}{2}
    \left\<\hat{\xi}(t)\,\hat{\xi}(t')\,+\,\hat{\xi}(t')\,\hat{\xi}(t)\right\>_{\rho_{b'}}\,
    =\, \frac{\hbar}{2} \sum_{j=1}^N \frac{c_j^2}{m_j\,\omega_j}
    \cos\{\omega_j (t-t')\}\,\coth\left(\frac{\beta \hbar \omega_j}{2}\right)\,.
\end{equation}

Now we introduce a response function \cite{ING98}
\begin{equation}\label{eq:response_fkt}
    \chi_{AB}(t)\, =\, \frac{i}{\hbar}\,\left\<[\hat{A}(t),
    \hat{B}(0)]\right\>_{\beta}\, \Theta(t)\,,
\end{equation}
where $\Theta(t)$ represents a step function. Then it can easily be
shown that $\chi_{pq}(t) = -\chi_{qp}(t) = M \dot{\chi}_{qq}(t)$ and
$\chi_{pp}(t) = -M^2 \ddot{\chi}_{qq}(t)$. For a later purpose it is
also necessary to discuss the time-reversal dynamics of $\hat{q}(t)$
in terms of $\hat{r}(t) := \hat{q}(-t)$ and its momentum $\hat{s}(t)
:= -\hat{p}(-t)$. We can then derive the corresponding quantum
Langevin equation \cite{HAE05}
\begin{equation}\label{eq:eq_of_motion1_for_x}
    M\,\ddot{\hat{r}}(t)\, +\,
    M \int_0^t d\tau\,\gamma(t-\tau)\,\dot{\hat{r}}(\tau)\,
    +\, M\,\omega_0^2\,\hat{r}(t)\; =\; \hat{\xi}_{-}(t)\,.
\end{equation}
While this is the same in form as Eq. (\ref{eq:eq_of_motion1}), the
two equations differ in the noise term in such a way that
$\hat{\xi}_{-}(t)$ is identical to $\hat{\xi}(t)$, however, with
replacement of $\hat{p}_j(0) \to -\hat{p}_j(0)$ in
(\ref{eq:damping_kernel1}). From Eq. (\ref{eq:response_fkt}) and the
stationarity relation $\<\hat{A}(t)\,\hat{B}(0)\>_{\beta} =
\<\hat{A}(0)\,\hat{B}(-t)\>_{\beta}$ \cite{ING98} it appears as well
that $\chi_{rr}(t) = -\chi_{qq}(t)$ and $\chi_{rs}(t) =
-\chi_{rp}(t) = -\chi_{qp}(t)$ and $\chi_{ss}(t) = -\chi_{pp}(t)$.
Applying the Laplace transform technique to Eqs.
(\ref{eq:eq_of_motion1}) and (\ref{eq:eq_of_motion1_for_x}),
respectively, we can finally obtain the exact expressions
\cite{ILK10}
\begin{subequations}
\begin{eqnarray}
    \hat{q}(t) &=& -\chi_{qp}(t)\,\hat{q} + \chi_{qq}(t)\,\hat{p} - \sum_{j} \{\chi_{q p_j}(t)\,\hat{x}_j - \chi_{q x_j}(t)\,\hat{p}_j\}\label{eq:q_time1}\\
    \hat{r}(t) &=& -\chi_{rp}(t)\,\hat{q} + \chi_{rr}(t)\,\hat{p} - \sum_{j} \{\chi_{r p_j}(t)\,\hat{x}_j - \chi_{r x_j}(t)\,\hat{p}_j\}\,,\label{eq:r_time1}
\end{eqnarray}
\end{subequations}
where the operators $\hat{q}, \hat{p}, \hat{x}_j$, and $\hat{p}_j$
represent the initial values $\hat{q}(0), \hat{p}(0), \hat{x}_j(0)$,
and $\hat{p}_j(0)$, respectively. Here we have
\begin{equation}\label{eq:system-bath-response-fkt1}
    \chi_{q x_j}(t)\, =\, \frac{1}{2\pi} \int_{-\infty}^{\infty}
    d\omega\,\tilde{\chi}_{q x_j}(\omega)\,e^{-i\omega t}
\end{equation}
with $\tilde{\chi}_{q x_j}(\omega) = c_j/\{m_j\,(\omega_j^2 -
\omega^2)\}\cdot\tilde{\chi}_{qq}(\omega)$, where the susceptibility
$\tilde{\chi}_{qq}(\omega)$ is the the Fourier-Laplace transform of
$\chi_{qq}(t)$. Also, $\chi_{r x_j}(t) = -\chi_{q x_j}(t)$, $\chi_{r
p_j}(t) = \chi_{q p_j}(t)$, and $\chi_{q p_j}(t) =
-m_j\,\dot{\chi}_{q x_j}(t)$.

It will also be useful later to introduce the well-known expressions
for the equilibrium fluctuations in terms of the susceptibility
$\tilde{\chi}_{qq}(\omega)$ such as \cite{FOR88}
\begin{subequations}
\begin{eqnarray}
    \<\hat{q}^2\>_{\beta} &=& \frac{\hbar}{\pi} \int_0^{\infty}
    d\omega\,\coth\left(\frac{\beta \hbar \omega}{2}\right)\,\text{Im}\{\tilde{\chi}_{qq}(\omega +
    i\,0^+)\}\label{eq:x_correlation1}\\
    \<\hat{p}^2\>_{\beta} &=& \frac{M^2 \hbar}{\pi} \int_0^{\infty}
    d\omega\,\omega^2\,\coth\left(\frac{\beta \hbar \omega}{2}\right)\,\text{Im}\{\tilde{\chi}_{qq}(\omega +
    i\,0^+)\}\,,\label{eq:x_dot_correlation1}
\end{eqnarray}
\end{subequations}
which can be derived from the fluctuation-dissipation theorem
\cite{CAL51}. For the Drude model (with a cut-off frequency
$\omega_d$ and a damping parameter $\gamma_o$), which is a prototype
for physically realistic damping, the equilibrium fluctuations are
explicitly given by \cite{KIM07}
\begin{eqnarray}
    \<\hat{q}^2\>_{\beta}^{(d)} &=& \frac{1}{M}
    \sum_{l=1}^3 \lambda_d^{(l)}\,\left\{\frac{1}{\beta \underline{\omega_l}}\,
    +\, \frac{\hbar}{\pi}\; \psi\left(\frac{\beta \hbar \underline{\omega_l}}{2 \pi}\right)\right\}\label{eq:x_drude}\\
    \<\hat{p}^2\>_{\beta}^{(d)} &=& -M
    \sum_{l=1}^3 \lambda_d^{(l)}\,\underline{\omega_l}^2\,\left\{\frac{1}{\beta \underline{\omega_l}}\,
    +\, \frac{\hbar}{\pi}\; \psi\left(\frac{\beta \hbar \underline{\omega_l}}{2 \pi}\right)\right\}\,,\label{eq:p_drude}
\end{eqnarray}
respectively, where the digamma function $\psi(y) =
d\,\ln\Gamma(y)/dy$ \cite{ABS74}, and $\underline{\omega_1} =
\Omega$, $\underline{\omega_2} = z_1$, $\underline{\omega_3} = z_2$,
and the coefficients
\begin{equation}\label{eq:coefficients}
    \lambda_d^{(1)}\; =\; \frac{z_1\,+\,z_2}{(\Omega\,-\,z_1) (z_2\,-\,\Omega)}\; ;\;
    \lambda_d^{(2)}\; =\; \frac{\Omega\,+\,z_2}{(z_1\,-\,\Omega)
    (z_2\,-\,z_1)}\; ;\;
    \lambda_d^{(3)}\; =\; \frac{\Omega\,+\,z_1}{(z_2\,-\,\Omega) (z_1\,-\,z_2)}\,.
\end{equation}
Here we have adopted, in place of $(\omega_0, \omega_d, \gamma_o)$,
the parameters $({\mathbf w}_0, \Omega, \gamma)$ through the
relations \cite{FOR06}
\begin{equation}\label{eq:parameter_change0}
    \omega_0^2\, :=\, {\mathbf w}_0^2\; \frac{\Omega}{\Omega\, +\, \gamma}\;\; ;\;\;
    \omega_d\, :=\, \Omega\, +\, \gamma\;\; ;\;\;
    \gamma_o\, :=\, \gamma\, \frac{\Omega\, (\Omega\, +\, \gamma)\,
    +\, {\mathbf w}_0^2}{(\Omega\, +\, \gamma)^2}\,,
\end{equation}
and then $z_1 = \gamma/2 + i {\mathbf w}_1$ and $z_2 = \gamma/2 - i
{\mathbf w}_1$ with ${\mathbf w}_1 = \sqrt{({\mathbf w}_0)^2 -
(\gamma/2)^2}$. From Eq. (\ref{eq:parameter_change0}) it also
follows that
\begin{equation}\label{eq:parameter-change1}
    \Omega^3 - \omega_d\,\Omega^2 + (\omega_0^2 + \gamma_0\,\omega_d)\,\Omega - \omega_d\,\omega_0^2\, =\,
    0\,,
\end{equation}
which will be used later. And we can then obtain for this damping
model the response function expressed as \cite{ILK10}
\begin{equation}\label{eq:response_fkt1}
    \chi_{qq}^{(d)}(t)\, =\, -\frac{1}{M} \frac{(z_1^2 - z_2^2)\,e^{-\Omega t} +
    (z_2^2 - \Omega^2)\,e^{-z_1 t} + (\Omega^2 - z_1^2)\,e^{-z_2 t}}{(\Omega - z_1) (z_1 - z_2) (z_2 -
    \Omega)}\,,
\end{equation}
which is real-valued and vanishes at $t = 0$ and $\infty$. It is
also interesting to note that this response function is
temperature-independent indeed, which was originally defined in
(\ref{eq:response_fkt}) as a function of temperature. In fact, the
susceptibility, defined as the Fourier-Laplace transform of the
response function $\chi_{qq}(t)$, is explicitly given by the
temperature-independent expression $\tilde{\chi}_{qq}(\omega) =
-1/\{M (\omega^2 + i \omega \tilde{\gamma}(\omega) - \omega_0^2)\}$
in terms of $\tilde{\gamma}(\omega)$ defined as the Fourier-Laplace
transform of the damping kernel $\gamma(t)$, which can easily be
obtained by applying the Laplace transform to Eq.
(\ref{eq:eq_of_motion1}) \cite{ING98}.

%%%%%%%%%%%%%%%%%%%%%%%%%%%%%%%%%%%%%%%%%%%%%%%%%%%%%%%%%%%%%%%%%%%%%%%%%%%%%
\section{Non-equilibrium process and its reduced density operator of the coupled oscillator}\label{sec:time-dpt-density-op}
%%%%%%%%%%%%%%%%%%%%%%%%%%%%%%%%%%%%%%%%%%%%%%%%%%%%%%%%%%%%%%%%%%%%%%%%%%%%%
%
Now we disturb the system of interest by varying its Hamiltonian
parameter with time, namely, either the spring constant $k(t)$ of
the coupled oscillator or its mass $M(t)$. Therefore, we should deal
with a time-dependent total system $\hat{{\mathcal H}}(t) =
\hat{{\mathcal H}}_s(t) + \hat{H}_b + \hat{H}_{sb}$ from now on,
where the time-dependent coupled oscillator $\hat{{\mathcal
H}}_s(t)$ is explicitly given by either
\begin{equation}\label{eq:time-dpt-hamiltonian1}
    \frac{\hat{p}^2}{2 M}\,+\,\frac{k(t)}{2}\,\hat{q}^2\, =\,
    \hat{H}_s\,+\,h_1(t)\,\hat{q}^2\, =: \,{}_{1}\hspace*{-.05cm}\hat{{\mathcal H}}_s(t)
\end{equation}
or
\begin{equation}\label{eq:time-dpt-hamiltonian2}
    \frac{\hat{p}^2}{2\,M(t)}\,+\,\frac{k_0}{2}\,\hat{q}^2\, =\,
    \hat{H}_s\,+\,h_2(t)\,\hat{p}^2\, =: \,{}_{2}\hspace*{-.05cm}\hat{{\mathcal H}}_s(t)\,.
\end{equation}
Here the initial values $k(0) = k_0$ and $M(0) = M$, and $h_1(t) =
\{k(t) - k_0\}/2$ and $h_2(t) = \{M - M(t)\}/\{2 M\cdot M(t)\}$. To
derive the (time-dependent) reduced density operator of the
oscillator $\hat{{\mathcal H}}_s(t)$, we first consider the equation
of motion for the density operator of the total system, which
explicitly reads as \cite{WEI08,ING98}
\begin{equation}\label{eq:density_fkt1}
    \hat{\rho}(t)\; =\; e^{-i t \hat{L}_0} \hat{\rho}(0)\,-\,i \int_0^t d\tau\,e^{-i (t - \tau) \hat{L}_0}
    \hat{L}_1(\tau)\,\hat{\rho}(\tau)\,.
\end{equation}
For a variation of the spring constant, the total Hamiltonian is
$\hat{{\mathcal H}}_1(t) = \hat{H}_0 + h_{1}(t)\,\hat{q}^2$, and the
corresponding Liouville operator $\hat{{\mathcal L}}^{(1)} =
\hat{L}_0 + \hat{L}^{(1)}_1$ satisfies
\begin{equation}\label{eq:density_fkt2}
    \hat{L}_0\,\hat{\rho}_1(t)\, =\, \frac{1}{\hbar}
    [\hat{H}_0, \hat{\rho}_1(t)]\;\;\; ;\;\;\;
    \hat{L}^{(1)}_1(\tau)\, \hat{\rho}_1(\tau)\, =\,
    \frac{1}{\hbar} [\hat{q}^2, \hat{\rho}_1(\tau)]\,h_1(\tau)\,.
\end{equation}
Here the Liouvillian $\hat{L}^{(1)}_1(\tau)$ surely corresponds to
$\hat{L}_1(\tau)$ in (\ref{eq:density_fkt1}). Likewise, for a
variation of the mass the total Hamiltonian is $\hat{{\mathcal
H}}_2(t) = \hat{H}_0 + h_{2}(t)\,\hat{p}^2$, and accordingly
$\hat{{\mathcal L}}^{(2)} = \hat{L}_0 + \hat{L}^{(2)}_1$ with
$\hat{L}^{(2)}_1 \leftarrow \hat{L}_1$ and
\begin{equation}\label{eq:density_fkt2_1}
    \hat{L}_0\,\hat{\rho}_2(t)\, =\, \frac{1}{\hbar}
    [\hat{H}_0, \hat{\rho}_2(t)]\;\;\; ;\;\;\;
    \hat{L}^{(2)}_1(\tau)\, \hat{\rho}_2(\tau)\, =\, \frac{1}{\hbar} [\hat{p}^2, \hat{\rho}_2(\tau)]\,h_2(\tau)\,.
\end{equation}

Now we attempt to obtain the density operator $\hat{\rho}_1(t)$ in
its explicit form. To this end, we mimic the technique applied for
the study of field-induced dynamics in the quantum Brownian
oscillator, discussed in \cite{ILK10}; we first substitute
(\ref{eq:density_fkt2}) into (\ref{eq:density_fkt1}), with
$\hat{\rho}_1(0) = \hat{\rho}_{\beta}$, and then make iterations for
$\hat{\rho}_1(\tau)$ in the integral. Then we can arrive at the
expression
\begin{eqnarray}\label{eq:density_fkt5_0}
    &&\textstyle \hat{\rho}_1(t)\, =\, \hat{\rho}_{\beta}\,-\,\frac{i}{\hbar} \int_0^t
    d\tau\,h_1(\tau)\,e^{-\frac{i}{\hbar} (t-\tau) \hat{H}_0}
    \left[\hat{q}^2, \hat{\rho}_{\beta}\right] e^{\frac{i}{\hbar} (t-\tau) \hat{H}_0}\n\\
    &&\textstyle +\,\left(-\frac{i}{\hbar}\right)^2
    \int_0^t d\tau\,h_1(\tau) \int_0^{\tau} d\tau'\,h_1(\tau')\,e^{-\frac{i}{\hbar} (t-\tau) \hat{H}_0}
    \left[\hat{q}^2, e^{-\frac{i}{\hbar} (\tau-\tau') \hat{H}_0}\left[\hat{q}^2,
    \hat{\rho}_{\beta}\right] e^{\frac{i}{\hbar} (\tau-\tau') \hat{H}_0}\right] e^{\frac{i}{\hbar} (t-\tau) \hat{H}_0}\n\\
    &&\textstyle +\,\left(-\frac{i}{\hbar}\right)^3 \int_0^t d\tau\,h_1(\tau) \int_0^{\tau} d\tau'\,h_1(\tau') \int_0^{\tau'}
    d\tau''\,h_1(\tau'')\,e^{-\frac{i}{\hbar} (t-\tau) \hat{H}_0}\n\\
    &&\textstyle \times\,\left[\hat{q}^2, e^{-\frac{i}{\hbar} (\tau-\tau') \hat{H}_0} \left[\hat{q}^2,
    e^{-\frac{i}{\hbar} (\tau'-\tau'') \hat{H}_0} \left[\hat{q}^2,
    \hat{\rho}_{\beta}\right] e^{\frac{i}{\hbar} (\tau'-\tau'') \hat{H}_0}\right]
    e^{\frac{i}{\hbar} (\tau-\tau') \hat{H}_0}\right] e^{\frac{i}{\hbar} (t-\tau) \hat{H}_0}\,+\,\cdots\,.
\end{eqnarray}
With the aid of $[\hat{\rho}_{\beta}, \hat{H}_0] = 0$, this equation
easily reduces to the expression in terms of $\hat{r}(t) =
e^{-\frac{i}{\hbar} t \hat{H}_0} \hat{q}\,e^{\frac{i}{\hbar} t
\hat{H}_0}$ as
\begin{eqnarray}\label{eq:density_fkt6_0}
    &&\textstyle \hat{\rho}_1(t)\, =\, \hat{\rho}_{\beta}\,+\,\frac{1}{i\hbar} \int_0^t
    d\tau\,h_1(\tau)\,\left[\hat{r}^2(t-\tau), \hat{\rho}_{\beta}\right]\,+\n\\
    &&\textstyle \left(\frac{1}{i\hbar}\right)^2 \int_0^t d\tau\,h_1(\tau)
    \int_0^{\tau} d\tau'\,h_1(\tau')\,\left[\hat{r}^2(t-\tau),
    \left[\hat{r}^2(t-\tau'), \hat{\rho}_{\beta}\right]\right]\,+\\
    &&\textstyle \left(\frac{1}{i\hbar}\right)^3
    \int_0^t d\tau\,h_1(\tau) \int_0^{\tau} d\tau'\,h_1(\tau') \int_0^{\tau'}
    d\tau''\,h_1(\tau'') \left[\hat{r}^2(t-\tau), \left[\hat{r}^2(t-\tau'), \left[\hat{r}^2(t-\tau''),
    \hat{\rho}_{\beta}\right]\right]\right]\,+\,\cdots\,,\n
\end{eqnarray}
where $\left[\hat{r}^2(t), \hat{\rho}_{\beta}\right] =
\hat{r}(t)\,\left[\hat{r}(t),
\hat{\rho}_{\beta}\right]\,+\,\left[\hat{r}(t),
\hat{\rho}_{\beta}\right]\,\hat{r}(t)$ with
\begin{equation}\label{eq:density_fkt7_0}
    \left[\hat{r}(t), \hat{\rho}_{\beta}\right]\, =\, {\textstyle-\chi_{rp}(t)
    \left[\hat{q}, \hat{\rho}_{\beta}\right]\,+\,\chi_{rr}(t) \left[\hat{p},
    \hat{\rho}_{\beta}\right]}\,-\,\sum_j {\textstyle\chi_{r p_j}(t) \left[\hat{x}_j,
    \hat{\rho}_{\beta}\right]}\,+\,\sum_j \textstyle\chi_{r x_j}(t) \left[\hat{p}_j,
    \hat{\rho}_{\beta}\right]\,,
\end{equation}
obtained directly from Eq. (\ref{eq:r_time1}). Likewise, we plug
(\ref{eq:density_fkt2_1}) into (\ref{eq:density_fkt1}) and then
apply the same technique as that used for (\ref{eq:density_fkt5_0}),
finally leading to the density operator $\hat{\rho}_2(t)$ in its
explicit form, identical to Eq.~(\ref{eq:density_fkt6_0}) but with
replacement of all $h_1(t) \to h_2(t)$ and all $\hat{r}^2(t) \to
\hat{s}^2(t)$ where $\hat{s}(t) = -e^{-\frac{i}{\hbar} t \hat{H}_0}
\hat{p}\,e^{\frac{i}{\hbar} t \hat{H}_0}$. Also, from
(\ref{eq:density_fkt7_0}) and $\hat{s}(t) = M \dot{\hat{r}}(t)$ we
have
\begin{equation}\label{eq:density_fkt7_1}
    \left[\hat{s}(t), \hat{\rho}_{\beta}\right]\, =\, {\textstyle-\chi_{sp}(t)
    \left[\hat{q}, \hat{\rho}_{\beta}\right]\,+\,\chi_{sr}(t) \left[\hat{p},
    \hat{\rho}_{\beta}\right]}\,-\,\sum_j {\textstyle\chi_{s p_j}(t) \left[\hat{x}_j,
    \hat{\rho}_{\beta}\right]}\,+\,\sum_j \textstyle\chi_{s x_j}(t) \left[\hat{p}_j,
    \hat{\rho}_{\beta}\right]\,.
\end{equation}

It is also instructive to rewrite Eq. (\ref{eq:density_fkt6_0}) as
its compact form
\begin{equation}\label{eq:reform_of_total_density_matrix1}
    \hat{\rho}_1(t)\,=\,T\,\exp\left\{\frac{1}{i\hbar} \int_0^t d\tau\,h_1(\tau)\,\hat{r}^2(t -
    \tau)\right\}\,\hat{\rho}_{\beta}\,\exp\left\{-\frac{1}{i\hbar} \int_0^t d\tau\,h_1(\tau)\,\hat{r}^2(t -
    \tau)\right\}\,,
\end{equation}
which is equivalent to
\begin{equation}\label{eq:reform_of_total_density_matrix11}
    \left(\tilde{T}\,\exp\left\{\frac{1}{i\hbar} \int_0^t d\tau\,h_1(\tau)\,\hat{r}^2(t -
    \tau)\right\}\right) \hat{\rho}_{\beta} \left(T\,\exp\left\{-\frac{1}{i\hbar} \int_0^t d\tau\,h_1(\tau)\,\hat{r}^2(t -
    \tau)\right\}\right)\,.
\end{equation}
The ordinary time ordering operator $T$ and its reverse time
ordering $\tilde{T}$ were introduced; in fact, $\hat{r}^2(t_1)$ and
$\hat{r}^2(t_2)$ do not commute at different times $t_1 \ne t_2$
[cf. Eqs.
(\ref{eq:r_squared_explicitly0})-(\ref{eq:explicit_commutator_relation1})].
Here we used the well-known operator identity \cite{GRI05}
\begin{equation}\label{eq:opeator_identity_Hadamard1}
    e^{\lambda \hat{B}}\,\hat{A}\,e^{-\lambda \hat{B}}\, =\, \hat{A}
    + \lambda \left[\hat{B}, \hat{A}\right] + \frac{\lambda^2}{2!} \left[\hat{B}, \left[\hat{B},
    \hat{A}\right]\right] + \frac{\lambda^3}{3!} \left[\hat{B}, \left[\hat{B}, \left[\hat{B}, \hat{A}\right]\right]\right] +
    \cdots\,.
\end{equation}
Let $\hat{r}^2(t) = \hat{\mathcal S}_1(t) + \hat{\mathcal B}_1(t) +
\hat{\mathcal C}_1(t)$, where explicitly
\begin{subequations}
\begin{eqnarray}
    \hat{\mathcal S}_1(t) &:=& \left\{\chi_{rp}(t)\,\hat{q} -
    \chi_{rr}(t)\,\hat{p}\right\}^2\label{eq:r_squared_explicitly0}\\
    \hat{\mathcal B}_1(t) &:=& \sum_{j,k} \left\{\chi_{r p_j}(t)\,\hat{x}_j -
    \chi_{r x_j}(t)\,\hat{p}_j\right\} \left\{\chi_{r p_k}(t)\,\hat{x}_k -
    \chi_{r x_k}(t)\,\hat{p}_k\right\}\\
    \hat{\mathcal C}_1(t) &:=& 2 \left\{\chi_{rp}(t)\,\hat{q} -
    \chi_{rr}(t)\,\hat{p}\right\} \sum_j \left\{\chi_{r p_j}(t)\,\hat{x}_j -
    \chi_{r x_j}(t)\,\hat{p}_j\right\}\,,\label{eq:r_squared_explicitly1}
\end{eqnarray}
\end{subequations}
and the three operators surely commute pairwise at any time $t$. On
the other hand, we have at $t_1 \ne t_2$
\begin{equation}\label{eq:commuting_system_bath1}
    \left[\hat{\mathcal S}_1(t_1), \hat{\mathcal B}_1(t_2)\right]\,=\,0\,,
\end{equation}
but all other commutators of the operators do not vanish indeed,
which are, respectively, in form of
\begin{subequations}
\begin{eqnarray}
    \left[\hat{\mathcal S}_1(t_1), \hat{\mathcal C}_1(t_2)\right] &=& \sum_j \left\{f_j^{(1)}(t_1,t_2)\,\hat{q} \hat{x}_j +
    f_j^{(2)}(t_1,t_2)\,\hat{q} \hat{p}_j + f_j^{(3)}(t_1,t_2)\,\hat{p} \hat{x}_j + f_j^{(4)}(t_1,t_2)\,\hat{p}
    \hat{p}_j\right\}\n\\\\
    \left[\hat{\mathcal B}_1(t_1), \hat{\mathcal C}_1(t_2)\right] &=& \sum_j \left\{g_j^{(1)}(t_1,t_2)\,\hat{q} \hat{x}_j +
    g_j^{(2)}(t_1,t_2)\,\hat{q} \hat{p}_j + g_j^{(3)}(t_1,t_2)\,\hat{p} \hat{x}_j + g_j^{(4)}(t_1,t_2)\,\hat{p} \hat{p}_j\right\}\n\\
\end{eqnarray}
\end{subequations}
as well as
\begin{subequations}
\begin{eqnarray}
    \left[\hat{\mathcal S}_1(t_1), \hat{\mathcal S}_1(t_2)\right] &=& \alpha_1(t_1,t_2)\,\hat{q}^2 + \alpha_2(t_1,t_2)\,\hat{p}^2 +
    \alpha_3(t_1,t_2)\,\left\{\hat{q} \hat{p} + \hat{p} \hat{q}\right\}\n\\\label{eq:explicit_commutator_relation0}\\
    \left[\hat{\mathcal B}_1(t_1), \hat{\mathcal B}_1(t_2)\right] &=& \sum_{j,k} \left\{\lambda_{jk}^{(1)}(t_1,t_2)\,\hat{x}_j \hat{x}_k +
    \lambda_{jk}^{(2)}(t_1,t_2)\,\hat{x}_j \hat{p}_k + \lambda_{jk}^{(3)}(t_1,t_2)\,\hat{p}_j \hat{x}_k +
    \lambda_{jk}^{(4)}(t_1,t_2)\,\hat{p}_j \hat{p}_k\right\}\n\\\\
    \left[\hat{\mathcal C}_1(t_1), \hat{\mathcal C}_1(t_2)\right] &=& \xi_1(t_1,t_2)\,\hat{q}^2 + \xi_2(t_1,t_2)\,\hat{p}^2 +
    \xi_3(t_1,t_2)\,\hat{q} \hat{p} + \xi_4(t_1,t_2)\,\hat{p} \hat{q} +\n\\
    && \sum_{j,k} \left\{\xi_{jk}^{(1)}(t_1,t_2)\,\hat{x}_j \hat{x}_k +
    \xi_{jk}^{(2)}(t_1,t_2)\,\hat{x}_j \hat{p}_k + \xi_{jk}^{(3)}(t_1,t_2)\,\hat{p}_j \hat{x}_k +
    \xi_{jk}^{(4)}(t_1,t_2)\,\hat{p}_j \hat{p}_k\right\}\,.\n\\\label{eq:explicit_commutator_relation1}
\end{eqnarray}
\end{subequations}
Here $f_j^{(n)}(t_1,t_2), g_j^{(n)}(t_1,t_2), \alpha_n(t_1,t_2),
\lambda_{jk}^{(n)}(t_1,t_2), \xi_n(t_1,t_2),
\xi_{jk}^{(n)}(t_1,t_2)$ are some scalar functions.

These non-commuting properties make it highly complicated to
explicitly carry out the transformation of the time-ordered
exponential operator $T\,e^{i\int\cdots\hat{r}^2(t-\tau)}$ in Eq.
(\ref{eq:reform_of_total_density_matrix11}), with the aid of Eqs.
(\ref{eq:commuting_system_bath1})-(\ref{eq:explicit_commutator_relation1})
and
(\ref{eq:exponential_identity1})-(\ref{eq:exponential_identity3}) as
well as the Zassenhaus formula and its dual, the
Baker-Campbell-Hausdorff formula \cite{WIL67,DON07}, into its
factorized form of
\begin{equation}
    e^{i f(\hat{q},\hat{p},t)}\cdot e^{i \sum_j (\cdots)}\cdot e^{i \sum_j
    g_j(\hat{x}_j,\hat{p}_j,t)}\,,
\end{equation}
where $(\cdots) = a_j(t)\,\hat{q}\hat{x}_j +
b_j(t)\,\hat{q}\hat{p}_j + c_j(t)\,\hat{p}\hat{x}_j +
d_j(t)\,\hat{p}\hat{p}_j$ in terms of the system-bath couplings
only. In fact, this transformation process is a critical step for
obtaining the reduced density operator $\hat{{\mathcal R}}_1(t) :=
\text{Tr}_b\,\hat{\rho}_1(t)$ of the coupled oscillator
${}_{1}\hspace*{-.05cm}\hat{{\mathcal H}}_s(t)$ in its closed form
from the total density operator $\hat{\rho}_1(t)$ in such a way
that, by the cyclic invariance of the trace,
\begin{eqnarray}\label{eq:reduced_density_operator_general_treatment1}
    \hat{{\mathcal R}}_1(t) &=& \mbox{Tr}_b\left\{e^{-i \sum_j g_j(\hat{x}_j,\hat{p}_j,t)}\, e^{-i \sum_j (\cdots)}\, e^{-i
    f(\hat{q},\hat{p},t)}\, \hat{\rho}_{\beta}\, e^{i f(\hat{q},\hat{p},t)}\, e^{i \sum_j (\cdots)}\, e^{i \sum_j g_j(\hat{x}_j,\hat{p}_j,t)}\right\}\n\\
    &=& \mbox{Tr}_b\left\{e^{-i \sum_j (\cdots)}\, e^{-i f(\hat{q},\hat{p},t)}\, \hat{\rho}_{\beta}\, e^{i f(\hat{q},\hat{p},t)}\,
    e^{i \sum_j (\cdots)}\,\right\}\,.
\end{eqnarray}
Here, $\text{Tr}_b$ denotes the partial trace for the bath alone.
And the initial equilibrium state $\hat{{\mathcal R}}(0) :=
\text{Tr}_b\,\hat{\rho}_{\beta}$ is defined as the reduced operator
of the canonical state $\hat{\rho}_{\beta}$ and explicitly given by
\cite{WEI08,GRA88}
\begin{equation}\label{eq:density_operator1}
    \<q|\hat{{\mathcal R}}(0)|q'\>\, =\, \frac{1}{\sqrt{2\pi \<\hat{q}^2\>_{\beta}}}\, \exp\left\{-\frac{(q + q')^2}{8\,\<\hat{q}^2\>_{\beta}} -
    \frac{\<\hat{p}^2\>_{\beta}\,(q - q')^2}{2 \hbar^2}\right\}\,,
\end{equation}
which holds true regardless of the system-bath coupling strengths.
Likewise $\hat{{\mathcal R}}_2(t) := \text{Tr}_b\,\hat{\rho}_2(t)$.

Consequently we now restrict our discussion for a closed form of the
reduced density operator $\hat{{\mathcal R}}_1(t)$ to the
weak-coupling limit, where $\chi_{r x_j}(t), \chi_{r p_j}(t) \to 0$
and so especially $\hat{\mathcal C}_1(t) \to 0$. From Eqs.
(\ref{eq:reform_of_total_density_matrix1}),
(\ref{eq:r_squared_explicitly1}) and
(\ref{eq:commuting_system_bath1})-(\ref{eq:reduced_density_operator_general_treatment1}),
it then follows that
\begin{equation}\label{eq:reform_of_total_density_matrix2}
    \hat{{\mathcal R}}_1^{(w)}(t)\, \approx\, T\,\exp\left\{\frac{1}{i\hbar} \int_0^t d\tau\,h_1(\tau)\,\hat{\mathcal S}_1(t -
    \tau)\right\}\,\hat{{\mathcal R}}(0)\,\exp\left\{-\frac{1}{i\hbar}
    \int_0^t d\tau\,h_1(\tau)\,\hat{\mathcal S}_1(t - \tau)\right\}\,.
\end{equation}
We stress here that this weak-coupling limit obviously differs from
an isolated system with identically vanishing coupling strengths
$(c_j \equiv 0)$; in fact, the response functions $\chi_{rp}(t)$ and
$\chi_{rr}(t)$ of $\hat{\mathcal S}_1(t)$ depend on the coupling
strengths already, as was discussed in Sect. \ref{sec:basics}. Also,
it is worthwhile to point out that Eq.
(\ref{eq:reform_of_total_density_matrix2}) can be regarded, by
construction, as a good short-time approximation to an exact
expression of the reduced density operator $\hat{{\mathcal
R}}_1(t)$. Further, as the response function $\chi^{(d)}_{qq}(t)$ in
(\ref{eq:response_fkt1}) and so the resulting quantities
$\{\chi^{(d)}_{r x_j}(t), \chi^{(d)}_{r p_j}(t)\}$ [cf.
(\ref{eq:system-bath-response-fkt1})] exponentially decay with time,
the contribution of $\hat{\mathcal C}_1(t)$ to the density operator
$\hat{{\mathcal R}}_1(t)$ may not be significantly non-negligible
with time $t$ large enough even in the strong-coupling limit, unless
$h_1(t)$ exponentially increases.

Let us now simplify the formal expression of the reduced density
operator $\hat{{\mathcal R}}_1^{(w)}(t)$ in
(\ref{eq:reform_of_total_density_matrix2}) by considering, with the
aid of (\ref{eq:opeator_identity_Hadamard1}), its expanded form such
as (\ref{eq:density_fkt6_0}); using Eq.
(\ref{eq:r_squared_explicitly0}) we can first obtain
\begin{equation}\label{eq:density_fkt8}
    \textstyle\<q|\text{Tr}_b \left[\hat{\mathcal S}_1(t), \hat{\rho}_{\beta}\right]|q'\>\, =\, \hat{{\mathcal A}}_{1}(t)\,\<q|\hat{{\mathcal R}}(0)|q'\>\,,
\end{equation}
where $\hat{{\mathcal A}}_{1}(t) := \chi_{rp}^2(t)\,\{q^2 - (q')^2\}
+ 2 i\hbar\,\chi_{rr}(t)\,\chi_{rp}(t)\,(q\partial_q +
\partial_{q'} q') + (i\hbar)^2 \chi_{rr}^2(t)\,(\partial_{q}^2 -
\partial_{q'}^2)$. Likewise, from (\ref{eq:density_fkt7_1}) and $\hat{s}^2(t) = \hat{\mathcal S}_2(t) +
\hat{\mathcal B}_2(t) + \hat{\mathcal C}_2(t)$ with $\hat{\mathcal
S}_2(t) := \{\chi_{sp}(t)\,\hat{q} - \chi_{sr}(t)\,\hat{p}\}^2$, we
can also have
\begin{equation}\label{eq:density_fkt8_1}
    \textstyle\<q|\text{Tr}_b \left[\hat{\mathcal S}_2(t), \hat{\rho}_{\beta}\right]|q'\>\, =\, \hat{{\mathcal A}}_{2}(t)\,\<q|\hat{{\mathcal R}}(0)|q'\>\,,
\end{equation}
in which $\hat{{\mathcal A}}_{2}(t) := \chi_{sp}^2(t)\,\{q^2 -
(q')^2\} + 2 i\hbar\,\chi_{sr}(t)\,\chi_{sp}(t)\,(q\partial_q +
\partial_{q'} q') + (i\hbar)^2 \chi_{sr}^2(t)\,(\partial_{q}^2 -
\partial_{q'}^2)$. Here we used $\text{Tr}_b\,([\hat{x}_j, \hat{\rho}_{\beta}]) = \text{Tr}_b\,([\hat{p}_j, \hat{\rho}_{\beta}]) =
0$. Similarly, $\text{Tr}_b\,(\hat{x}_j\,[\hat{p},
\hat{\rho}_{\beta}]) = \text{Tr}_b\,(\hat{p}_j\,[\hat{q},
\hat{\rho}_{\beta}]) = \text{Tr}_b\,(\hat{p}_j\,[\hat{p},
\hat{\rho}_{\beta}]) = 0$. And
\begin{eqnarray}\label{eq:density_fkt10}
    \hspace*{-.5cm}\<q|\,\text{Tr}_b \left\{\hat{q}, \left[\hat{q},
    \hat{\rho}_{\beta}\right]\right\}_+|q'\> &=& \left\{q^2 - (q')^2\right\}\,\<q|\hat{{\mathcal
    R}}(0)|q'\>\n\\
    \hspace*{-.5cm}\<q|\,\text{Tr}_b\left(\left\{\hat{q}, \left[\hat{p},
    \hat{\rho}_{\beta}\right]\right\}_+ + \left\{\hat{p}, \left[\hat{q},
    \hat{\rho}_{\beta}\right]\right\}_+\right)|q'\> &=&
    -2 i \hbar \left(q \partial_q + \partial_{q'} q'\right)\,\<q|\hat{{\mathcal R}}(0)|q'\>\n\\
    \hspace*{-.5cm}\<q|\,\text{Tr}_b \left\{\hat{p}, \left[\hat{p},
    \hat{\rho}_{\beta}\right]\right\}_+|q'\> &=&
    \left(i \hbar\right)^2 \left(\partial_q^2 - \partial_{q'}^2\right)\,\<q|\hat{{\mathcal
    R}}(0)|q'\>\,,
\end{eqnarray}
where the anticommutator $\{\hat{A}, \hat{B}\}_+ = \hat{A} \hat{B} +
\hat{B} \hat{A}$.

With the aid of Eqs.
(\ref{eq:density_fkt8})-(\ref{eq:density_fkt10}), we can next obtain
\begin{subequations}
\begin{eqnarray}
    \<q|\text{Tr}_b \left[\hat{\mathcal S}_1(t),
    \left[\hat{\mathcal S}_1(\tau), \hat{\rho}_{\beta}\right]\right]|q'\> &=&
    \textstyle\hat{{\mathcal A}}_{1}(t)\, \hat{{\mathcal A}}_{1}(\tau)\, \<q|\hat{{\mathcal R}}(0)|q'\>\label{eq:density_fkt13}\\
    \<q|\text{Tr}_b \left[\hat{\mathcal S}_2(t),
    \left[\hat{\mathcal S}_2(\tau), \hat{\rho}_{\beta}\right]\right]|q'\> &=&
    \textstyle\hat{{\mathcal A}}_{2}(t)\, \hat{{\mathcal A}}_{2}(\tau)\, \<q|\hat{{\mathcal R}}(0)|q'\>\,.\label{eq:density_fkt13_1}
\end{eqnarray}
\end{subequations}
Along the same line, after making a lengthy calculation, we can also
arrive at the expressions
\begin{subequations}
\begin{eqnarray}
    \hspace*{-1cm}\<q|\text{Tr}_b \left[\hat{\mathcal S}_1(t), \left[\hat{\mathcal S}_1(\tau),
    \left[\hat{\mathcal S}_1(\tau'), \hat{\rho}_{\beta}\right]\right]\right]|q'\> &=& \hat{{\mathcal A}}_{1}(t)\, \hat{{\mathcal A}}_{1}(\tau)\,
    \hat{{\mathcal A}}_{1}(\tau')\, \<q|\hat{{\mathcal R}}(0)|q'\>\label{eq:density_fkt14}\\
    \hspace*{-1cm}\<q|\text{Tr}_b \left[\hat{\mathcal S}_2(t), \left[\hat{\mathcal S}_2(\tau),
    \left[\hat{\mathcal S}_2(\tau'), \hat{\rho}_{\beta}\right]\right]\right]|q'\> &=& \hat{{\mathcal A}}_{2}(t)\, \hat{{\mathcal A}}_{2}(\tau)\,
    \hat{{\mathcal A}}_{2}(\tau')\, \<q|\hat{{\mathcal R}}(0)|q'\>\,.\label{eq:density_fkt14_1}
\end{eqnarray}
\end{subequations}
Based on Eqs. (\ref{eq:density_fkt6_0}),
(\ref{eq:density_fkt8})-(\ref{eq:density_fkt8_1}), and
(\ref{eq:density_fkt13})-(\ref{eq:density_fkt14_1}) we can finally
find the matrix elements of the reduced density operator of the
coupled oscillator ${}_{1}\hspace*{-.05cm}\hat{{\mathcal H}}_s(t)$
as
\begin{equation}\label{eq:density_fkt16}
    \<q|\hat{{\mathcal R}}^{(w)}_1(t)|q'\>\, =\, T\,e^{-\frac{i}{\hbar} \hat{J}_1(t)}\, \<q|\hat{{\mathcal R}}(0)|q'\>\,,
\end{equation}
where $\hat{J}_1(t) = \int_0^t d\tau\,h_1(\tau)\,\hat{{\mathcal
A}}_{1}(t-\tau)$ represents the time-evolution action. Likewise, the
density operator of the coupled oscillator
${}_{2}\hspace*{-.05cm}\hat{{\mathcal H}}_s(t)$ can be obtained as
\begin{equation}\label{eq:density_fkt16_0}
    \<q|\hat{{\mathcal R}}^{(w)}_2(t)|q'\>\, =\, T\,e^{-\frac{i}{\hbar} \hat{J}_2(t)}\, \<q|\hat{{\mathcal
    R}}(0)|q'\>\,,
\end{equation}
where $\hat{J}_2(t) = \int_0^t d\tau\,h_2(\tau)\,\hat{{\mathcal
A}}_{2}(t-\tau)$.

Now we see from Eq. (\ref{eq:explicit_commutator_relation0}) that
$\hat{\mathcal A}_{1}(t_1)$ and $\hat{\mathcal A}_{1}(t_2)$ at $t_1
\ne t_2$ are not commuting and accordingly it is non-trivial to
directly deal with the time-ordered exponential operator in
(\ref{eq:density_fkt16}). Therefore, we need to introduce the
exponential operator identity derived in \cite{LAM98}
\begin{equation}\label{eq:exponential_identity1}
    T\,\exp\left\{\int_0^t d\tau\,\hat{\mathcal{B}}_t(\tau)\right\}\, =\,
    \exp\left\{\sum_{n=1}^{\infty} \hat{K}_n(t)\right\}\,,
\end{equation}
where $\hat{\mathcal{B}}_t(\tau) := h_1(\tau)\,\hat{{\mathcal
A}}_{1}(t-\tau)$, and the low-order terms are explicitly given by
\begin{eqnarray}\label{eq:exponential_identity2}
    \hat{K}_1(t)\, =\, \hat{C}_1(t) \;\;\;&;&\;\;\; \hat{K}_2(t)\, =\, \frac{1}{2}\,\hat{C}_2(t)\n\\
    \hat{K}_3(t)\, =\, \frac{1}{3}\,\hat{C}_3(t) + \frac{1}{12}\,\left[\hat{C}_2(t), \hat{C}_1(t)\right] \;\;\;&;&\;\;\;
    \hat{K}_4(t)\, =\, \frac{1}{4}\,\hat{C}_4(t) + \frac{1}{12}\,\left[\hat{C}_3(t),
    \hat{C}_1(t)\right]\,.\n\\
\end{eqnarray}
Here the commutators
\begin{equation}\label{eq:exponential_identity3}
    \hat{C}_n(t)\, =\, \int_0^t d\tau_1 \int_0^{\tau_1} d\tau_2 \cdots \int_0^{\tau_{n-1}} d\tau_n\,
    \left[\hat{\mathcal{B}}_t(\tau_1), \left[\hat{\mathcal{B}}_t(\tau_2),
    \left[\cdots, \left[\hat{\mathcal{B}}_t(\tau_{n-1}), \hat{\mathcal{B}}_t(\tau_{n})\right] \cdots\right]\right]\right]
\end{equation}
with $\hat{C}_1(t) = \int_0^t d\tau\,\hat{\mathcal{B}}_t(\tau)$. In
fact, the operators $\hat{K}_n(t)$ for all $n$ can be evaluated
exactly.

Let us simplify the commutators $\hat{C}_n(t)$ to derive the closed
expression of $\<q|\hat{{\mathcal R}}^{(w)}_1(t)|q'\>$ in
(\ref{eq:density_fkt16}). First let $y := q+q'$ and $z := q-q'$, and
so $\partial_q + \partial_{q'} = 2
\partial_y$ and $\partial_q -
\partial_{q'} = 2
\partial_z$. Then it easily appears that $\hat{{\mathcal A}}_{1}(t) = \{\chi_{rp}(t)\,y +
2 i\hbar\,\chi_{rr}(t)\,\partial_z\} \{\chi_{rp}(t)\,z + 2
i\hbar\,\chi_{rr}(t)\,\partial_y\}$. This allows us to finally
obtain
\begin{equation}\label{eq:density_op6}
    \hat{C}_n(t)\, =\, \frac{a_n(t)}{i \hbar}\,y z + b_{n}(t)\,(y \partial_y + \partial_z z) + i \hbar\,c_{n}(t)\,\partial_y \partial_z\,,
\end{equation}
where $a_n(t), b_n(t)$ and $c_n(t) \in {\mathbb R}$, and explicitly
given by
\begin{eqnarray}\label{eq:density_op61}
    a_{2m-1}(t)\, =\, 2^{3m-3}\,I_{2m-1,1}(t,t)\;\;\; &;&\;\;\; a_{2m}(t)\, =\,
    2^{3m-1}\,I_{2m,1}(t,t)\n\\
    b_{2m-1}(t)\, =\, 2^{3m-2}\,I_{2m-1,2}(t,t)\;\;\;
    &;&\;\;\; b_{2m}(t)\, =\, 2^{3m-1}\,I_{2m,2}(t,t)\n\\
    c_{2m-1}(t)\, =\, 2^{3m-1}\,I_{2m-1,3}(t,t)\;\;\; &;&\;\;\; c_{2m}(t)\, =\, 2^{3m+1}\,I_{2m,3}(t,t)
\end{eqnarray}
with $m = 1,2,3, \cdots$. Here
\begin{eqnarray}\label{eq:density_op7}
    &\textstyle I_{11}(t,u)\, :=\, \int_0^u
    d\tau\,h_1(\tau)\,\chi_{rp}^2(t-\tau)\;\;\; ;\;\;\; I_{12}(t,u)\, :=\, \int_0^u d\tau\,h_1(\tau)\,\chi_{rr}(t-\tau)\,\chi_{rp}(t-\tau)&\n\\
    &\textstyle I_{13}(t,u)\, :=\, \int_0^u d\tau\,h_1(\tau)\,\chi_{rr}^2(t-\tau)\,.&
\end{eqnarray}
For $n \geq 2$, we have
\begin{equation}\label{eq:density_op70}
    I_{n2}(t,u)\, :=\, \textstyle\int_0^u d\tau\,h_1(\tau)\,\{\chi_{rr}^2(t-\tau)\,I_{n-1,1}(t,\tau) -
    \chi_{rp}^2(t-\tau)\,I_{n-1,3}(t,\tau)\}\,,
\end{equation}
and
\begin{eqnarray}\label{eq:density_op72}
    I_{2m,1}(t,u) &:=& \textstyle\int_0^u d\tau\,h_1(\tau)\,\{\chi_{rr}(t-\tau)\,\chi_{rp}(t-\tau)\,I_{2m-1,1}(t,\tau) - \chi_{rp}^2(t-\tau)\,I_{2m-1,2}(t,\tau)\}\n\\
    I_{2m,3}(t,u) &:=& \textstyle\int_0^u d\tau\,h_1(\tau)\,\{\chi_{rr}^2(t-\tau)\,I_{2m-1,2}(t,\tau) -
    \chi_{rr}(t-\tau)\,\chi_{rp}(t-\tau)\,I_{2m-1,3}(t,\tau)\}\n\\
\end{eqnarray}
as well as
\begin{eqnarray}\label{eq:density_op74}
    I_{2m+1,1}(t,u) &:=& \textstyle\int_0^u d\tau\,h_1(\tau)\,\{2\,\chi_{rr}(t-\tau)\,\chi_{rp}(t-\tau)\,I_{2m,1}(t,\tau) - \chi_{rp}^2(t-\tau)\,I_{2m,2}(t,\tau)\}\n\\
    I_{2m+1,3}(t,u) &:=& \textstyle\int_0^u d\tau\,h_1(\tau)\,\{\chi_{rr}^2(t-\tau)\,I_{2m,2}(t,\tau) -
    2\,\chi_{rr}(t-\tau)\,\chi_{rp}(t-\tau)\,I_{2m,3}(t,\tau)\}\,.\n\\
\end{eqnarray}
Here we employed the commutators in (\ref{eq:commutators_1}) with
the replacement of $\hat{M} \to y\partial_y +
\partial_z z$ and $\hat{A} \to yz$ and $\hat{B} \to
\partial_y \partial_z$, which immediately leads to $\alpha \to 2, m \to
-1$ and $r \to 0$. With the aid of Eqs.
(\ref{eq:exponential_identity1})-(\ref{eq:density_op6}) we can then
rewrite the time-evolution in (\ref{eq:density_fkt16}) as the
unitary operator
\begin{equation}\label{eq:density_op75}
    T\,\exp{\left\{-\frac{i}{\hbar}\,\hat{J}_1(t)\right\}}\, =\,
    \exp\left\{\frac{\tilde{a}_1(t)}{i \hbar}\,y z + \tilde{b}_1(t)\,(y \partial_y + \partial_z z) + i \hbar\,\tilde{c}_1(t)\,\partial_y
    \partial_z\right\}\,,
\end{equation}
where the real-valued coefficients
\begin{subequations}
\begin{eqnarray}
    \tilde{a}_1(t) &:=& a_1(t) + \frac{a_2(t)}{2} + \frac{a_3(t)}{3} + \frac{1}{6}
    \left\{a_1(t)\,b_2(t) - b_1(t)\,a_2(t)\right\} +\n\\
    && \frac{a_4(t)}{4} + \frac{1}{6} \left\{a_1(t)\,b_3(t) - b_1(t)\,a_3(t)\right\} + \cdots\label{eq:density_op761}\\
    \tilde{b}_1(t) &:=& b_1(t) + \frac{b_2(t)}{2} + \frac{b_3(t)}{3} + \frac{1}{12}
    \left\{a_1(t)\,c_2(t) - c_1(t)\,a_2(t)\right\} +\n\\
    && \frac{c_4(t)}{4} + \frac{1}{12} \left\{a_1(t)\,c_3(t) - c_1(t)\,a_3(t)\right\} + \cdots\label{eq:density_op762}\\
    \tilde{c}_1(t) &:=& c_1(t) + \frac{c_2(t)}{2} + \frac{c_3(t)}{3} - \frac{1}{6}
    \left\{c_1(t)\,b_2(t) - b_1(t)\,c_2(t)\right\} +\n\\
    && \frac{c_4(t)}{4} - \frac{1}{6} \left\{c_1(t)\,b_3(t) - b_1(t)\,c_3(t)\right\} + \cdots\label{eq:density_op763}
\end{eqnarray}
\end{subequations}
(in fact, all higher-order terms can be determined exactly).
Likewise, coefficients $\tilde{a}_2(t), \tilde{b}_2(t)$ and
$\tilde{c}_2(t)$ pertaining to Eq. (\ref{eq:density_fkt16_0}) can
also be introduced, which are identical to their counterparts in
(\ref{eq:density_op761})-(\ref{eq:density_op763}), respectively,
however obtained from the replacement of $h_1 \to h_2$ and
$\chi_{rr} \to \chi_{sr}$ and $\chi_{rp} \to \chi_{sp}$ in
(\ref{eq:density_op7})-(\ref{eq:density_op74}).

To further proceed with (\ref{eq:density_op75}), we apply another
operator identity, derived in \cite{MIT84}, given by
\begin{equation}\label{eq:density_op8}
    \exp\left\{\lambda\,(\hat{M} + \mu \hat{A} + \nu \hat{B})\right\}\; =\;
    \exp\left(\kappa \hat{M}\right)\,\exp\left(f e^{-\alpha \kappa} \hat{A}\right)\,\exp\left(g \hat{B}\right)\,\exp\left(d\right)
\end{equation}
where $\lambda, \mu, \nu$ are arbitrary complex numbers, and the
product $\mu \nu$ is assumed to be real-valued. Here the operators
$\hat{M}, \hat{A}, \hat{B}$ satisfy the commutator relations
\begin{equation}\label{eq:commutators_1}
    \left[\hat{A}, \hat{M}\right]\, =\, -\alpha \hat{A}\;\;\; ;\;\;\; \left[\hat{B}, \hat{M}\right]\, =\, \alpha
    \hat{B}\;\;\; ;\;\;\; \left[\hat{A}, \hat{B}\right]\, =\, m \hat{M} + r
\end{equation}
where $\alpha, m, r \in {\mathbb R}$. And numerical functions $f =
\mu X$ and $g = \nu X$ where $X = (\tan \lambda D)/\{D - (\alpha/2)
\tan \lambda D\}$ with $D^2 = -(\mu \nu m + \alpha/2) (\alpha/2)$,
and $\kappa = -(2/\alpha) \ln\left\{\cos \lambda D - (\alpha/2D)
\sin \lambda D\right\}$ and $d = (\kappa - \lambda) r/m$.

After making a lengthy calculation with the aid of Eq.
(\ref{eq:density_op8}), every single step of which is provided in
detail in Appendix, we can finally arrive at the closed expression
\begin{equation}\label{eq:density_matrix1}
    \<q|\hat{{\mathcal R}}^{(w)}_1(t)|q'\>\, =\,
    \sqrt{\frac{B_1(t)}{2\pi\,\<\hat{q}^2\>_{\beta}}}\, \exp\left(-B_{1}(t) \left\{\frac{(q +
    q')^2}{8\,\<\hat{q}^2\>_{\beta}} + \frac{\<\hat{p}^2\>_{\beta}\,(q -
    q')^2}{2\,\hbar^2}\right\} + i {\Phi}_{1}(t)\,\frac{q^2 - q'^2}{\<\hat{q}^2\>_{\beta}}\right)
\end{equation}
where the two dimensionless parameters
\begin{eqnarray}
    B_{1}(t) &:=& \left(\frac{D(t)}{D(t) \cos\{\tilde{b}_1(t)\,D(t)\} - \sin\{\tilde{b}_1(t)\,D(t)\}}\right)^2\,\left\{1 - \frac{g^2(t)}{4 \hbar^2}
    \frac{\<\hat{p}^2\>_{\beta}}{\<\hat{q}^2\>_{\beta}}\right\}^{-1} \in {\mathbb R}^+\n\\\label{eq:density_matrix21}\\
    \Phi_{1}(t) &:=& \frac{\<\hat{p}^2\>_{\beta}}{4 i \hbar^2}\, g(t)\, B_{1}(t)\, -\, i \<\hat{q}^2\>_{\beta}\, f(t)\,\in {\mathbb
    R}\label{eq:density_matrix22}
\end{eqnarray}
in terms of the coefficients $\tilde{a}_1(t), \tilde{b}_1(t)$ and
$\tilde{c}_1(t)$ in
(\ref{eq:density_op761})-(\ref{eq:density_op763}). Here, the
parameter $D(t) = \pm
\{\tilde{a}_1(t)\,\tilde{c}_1(t)/\tilde{b}_1^2(t) - 1\}^{1/2} \in
{\mathbb R}$ or $i {\mathbb R}$ as given in Appendix and so $f(t)$
and $g(t) \in i {\mathbb R}$ in (\ref{eq:density_op16_1}). As shown,
the time-dependency of the reduced density operator in
(\ref{eq:density_matrix1}) consists entirely in $B_{1}(t)$ and
$\Phi_{1}(t)$. Figs. \ref{fig:fig1}-\ref{fig:fig2} demonstrate their
behaviors versus time for, e.g., $k(t) = k_0\,(2-e^{-t})$ within the
Drude damping model. Applying exactly the same technique, we can
also derive the reduced density operator $\<q|\hat{{\mathcal
R}}^{(w)}_2(t)|q'\>$ in closed form, which is in fact identical to
Eq. (\ref{eq:density_matrix1}) but with replacement of $B_1(t),
\Phi_1(t) \to B_2(t), \Phi_2(t)$ in terms of $\tilde{a}_2(t),
\tilde{b}_2(t)$ and $\tilde{c}_2(t)$ for $f(t), g(t)$ and $D(t)$
therein. The normalization $\text{Tr}\,\hat{\mathcal R}^{(w)}_1(t) =
\text{Tr}\,\hat{\mathcal R}^{(w)}_2(t) = 1$ can easily be verified
with the aid of the identity \cite{ABS74}
\begin{equation}\label{eq:integral1}
    \int_{-\infty}^{\infty} dx\, e^{-(a x^2 + 2 b x)}\;
    =\; \sqrt{\frac{\pi}{a}}\, e^{\frac{b^2}{a}}\,.
\end{equation}
It then follows that $\<\hat{q}\>_{\rho_1(t)} \equiv
\<\hat{q}\>_{{\mathcal R}^{(w)}_1(t)} = 0$ and
$\<\hat{p}\>_{\rho_1(t)} \equiv \<\hat{p}\>_{{\mathcal
R}^{(w)}_1(t)} = 0$, and
\begin{equation}\label{eq:expectation_values2}
    \<\hat{q}^2\>_{\rho_1(t)}\, =\, \frac{1}{B_{1}(t)}\, \<\hat{q}^2\>_{\beta}\;\;\; ;\;\;\;
    \<\hat{p}^2\>_{\rho_1(t)}\, =\, B_{1}(t)\, \<\hat{p}^2\>_{\beta}\, +\, \frac{4 \hbar^2\,\Phi_{1}^2(t)}{B_{1}(t)}\, \frac{1}{\<\hat{q}^2\>_{\beta}}\,.
\end{equation}
From this, the instantaneous uncertainty relation also follows as
\begin{equation}\label{eq:uncertainty1}
    (\Delta q)^2_{\rho_1(t)}\, (\Delta p)^2_{\rho_1(t)}\; =\;
    \<\hat{q}^2\>_{\beta}\, \<\hat{p}^2\>_{\beta}\, +\, \frac{4 \hbar^2\,\Phi_{1}^2(t)}{B_{1}^2(t)}\,.
\end{equation}
Then the instantaneous internal energy of the coupled oscillator
reads as
\begin{equation}\label{eq:hamiltonian_internal_energy1}
    {}_{1}\hspace*{-.01cm}{\mathcal U}_s(t)\, =\, \<{}_{1}\hspace*{-.05cm}\hat{{\mathcal H}}_s(t)\>_{{\mathcal R}^{(w)}_1(t)}\, =\,
    \frac{\<\hat{p}^2\>_{\rho_1(t)}}{2 M}\, +\,
    \frac{k(t)}{2}\,\<\hat{q}^2\>_{\rho_1(t)}\,.
\end{equation}
Along the same line, the expectation values for the density operator
$\hat{{\mathcal R}}^{(w)}_2(t)$ easily appear, respectively, as the
counterparts to those in Eqs. (\ref{eq:expectation_values2}) and
(\ref{eq:uncertainty1}) in terms of $B_2(t)$ and $\Phi_2(t)$, and so
the instantaneous internal energy ${}_{2}\hspace*{-.01cm}{\mathcal
U}_s(t) = \<{}_{2}\hspace*{-.05cm}\hat{{\mathcal
H}}_s(t)\>_{{\mathcal R}^{(w)}_2(t)}$ will immediately follow as
well.

Comments deserve here. The compact form of the density operator
$\hat{{\mathcal R}}^{(w)}_1(t)$ in (\ref{eq:density_matrix1}), valid
for an arbitrary variation of the spring constant $k(t)$, was
clearly derived for the special initial condition $\hat{{\mathcal
R}}(0)$ in (\ref{eq:density_operator1}), or equivalently, the
canonical thermal equilibrium state $\hat{\rho}_{\beta}$ of the
total system $\hat{H}_0$ with $[\hat{\rho}_{\beta}, \hat{H}_0] = 0$.
This then gave rise to the significant simplification in form in the
step from (\ref{eq:density_fkt5_0}) to (\ref{eq:density_fkt6_0}),
which subsequently led, with the useful relations in
(\ref{eq:density_fkt10}), to Eq. (\ref{eq:density_fkt16}) and
finally Eq. (\ref{eq:density_matrix1}). In the general case of the
initial condition, on the other hand, it is mathematically not
straightforward to obtain an explicit form of the density operator
$\hat{{\mathcal R}}^{(w)}_1(t)$. Clearly, the time-dependent
coefficients $\{\tilde{a}_1(t), \tilde{b}_1(t), \tilde{c}_1(t)\}$ in
(\ref{eq:density_op761})-(\ref{eq:density_op763}) are fundamental
ingredients to the time-evolution operator in
(\ref{eq:density_op75}) and so the reduced density operator
$\hat{{\mathcal R}}^{(w)}_1(t)$. Then, as shown in
(\ref{eq:density_op61})-(\ref{eq:density_op74}), the coefficients
are expressed in terms of the parameter $h_1(t)$ representing the
variation of the spring constant as well as the response functions
$\chi_{rr}(t)$ and $\chi_{rp}(t)$, defined as the average values
with respect to the initial condition $\hat{\rho}_{\beta}$ but
temperature-independent indeed [cf. (\ref{eq:response_fkt1})] and
reflecting the characteristics of the bath coupled to the oscillator
in consideration.

It may also be worthwhile to point out that Zerbe and H\"{a}nggi
derived a master equation for the reduced density operator
$\hat{{\mathcal R}}_1(t)$, however, restricted to i) the periodic
potential, $h_1(t)\,\hat{q}^2 \to (m/2)\cdot\epsilon\,\cos (\Omega t
+ \varphi)\,\hat{q}^2$; ii) the Ohimic damping; iii) the initial
state of the total system given by an uncoupled one $\hat{\rho}_0 =
\hat{\rho}_s \otimes \hat{\rho}_b$ \cite{HAE95}, whereas this is
obviously not the case in our study. Accordingly, the initial state
$\hat{\rho}_0$ cannot represent a thermal equilibrium of the coupled
total system (oscillator plus bath), which is necessary for the
discussion of the Clausius inequality in Sect.
\ref{sec:thermodynamics}. Further, in a damping model without
cut-off frequency (such as the Ohmic), which is not physically
realistic, the validity of the second law in the quantum Brownian
oscillator may not be guaranteed \cite{KIM07,KIM06}.

%%%%%%%%%%%%%%%%%%%%%%%%%%%%%%%%%%%%%%%%%%%%%%%%%%%%%%%%%%%%%%%%%%%%%%%%%%%%%
\section{Quasi-static process and its reduced density operator of the coupled oscillator}\label{sec:instantaneous-eigenvalue}
%%%%%%%%%%%%%%%%%%%%%%%%%%%%%%%%%%%%%%%%%%%%%%%%%%%%%%%%%%%%%%%%%%%%%%%%%%%%%
%
For comparison with the above non-equilibrium processes, we discuss
the corresponding quasi-static processes. Here the system of
interest undergoes change infinitely slowly and so remains in
equilibrium exactly in form of Eq. (\ref{eq:density_operator1}) in
every single step such that for any spring constant $k$,
\begin{equation}\label{eq:density_operator10}
    \<q|\hat{R}_{\text{\sc eq}}(k)|q'\>\, =\, \frac{1}{\sqrt{2\pi \<\hat{q}^2\>_{\beta}(k)}}\, \exp\left\{-\frac{(q + q')^2}{8\,\<\hat{q}^2\>_{\beta}(k)} -
    \frac{\<\hat{p}^2\>_{\beta}(k)\cdot(q - q')^2}{2
    \hbar^2}\right\}\,,
\end{equation}
(valid for an arbitrary system-bath coupling strength indeed), where
the initial values $\<\hat{q}^2\>_{\beta}(k_0) =
\<\hat{q}^2\>_{\beta}$ and $\<\hat{p}^2\>_{\beta}(k_0) =
\<\hat{p}^2\>_{\beta}$. Apparently, this density matrix looks
different from $\<q|\hat{{\mathcal R}}^{(w)}_{1}(t)|q'\>$ in
(\ref{eq:density_matrix1}), and in general not in form of a
canonical thermal state $\propto
e^{-\beta\,{}_{1}\hspace*{-.05cm}\hat{{\mathcal H}}_s(t)}$
\cite{KIM10}. Eq. (\ref{eq:density_matrix1}), however, reduces to
its quasi-static counterpart in (\ref{eq:density_operator10}) indeed
if $\dot{k}(t) \to 0$ at every single moment: As demonstrated in
Fig. \ref{fig:fig2} for the parameter $\Phi_1(t)$ of
(\ref{eq:density_matrix1}), where $k(t) = k_0\,(2 - e^{-t})$ and so
$\dot{k}(\infty) \to 0$, we have $\dot{\Phi}_1(\infty) \to 0$. From
this and the initial value $\Phi_1(0) = 0$, it must follow that if
$\dot{k}(t)$ remains infinitesimally small at every single moment,
then $\Phi_1(t) \to 0$ always. This immediately leads to
$\<\hat{p}^2\>_{\rho_1(t)} \to B_{1}(t)\,\<\hat{p}^2\>_{\beta}$ in
(\ref{eq:expectation_values2}), and then $\<\hat{q}^2\>_{\rho_1(t)}
= \<\hat{q}^2\>_{\beta}/B_{1}(t) \to \<\hat{q}^2\>_{\beta}(k)$ as
well as $\<\hat{p}^2\>_{\rho_1(t)} \to \<\hat{p}^2\>_{\beta}(k)$ in
(\ref{eq:density_matrix1}). As a result, we can arrive at Eq.
(\ref{eq:density_operator10}).

Consequently, without any harm we can straightforwardly adopt here,
with $k_0 \to k$, all results for the initial equilibrium state
$\hat{R}_{\text{\sc eq}}(k_0) = \hat{{\mathcal R}}(0)$ obtained in
\cite{KIM10}; we can introduce an {\em uncoupled} effective
oscillator
\begin{equation}\label{eq:hamiltonian_reversible1}
    \hat{H}_{\text{\sc eff}}^{\star}(k_0)\; =\; \frac{\hat{p}^2}{2\,M_{\text{\sc eff}}^{\star}(k_0)}\, +\,
    \frac{k_{\text{\sc eff}}^{\star}(k_0)}{2}\,\hat{q}^2
\end{equation}
in the same state $\hat{R}_{\text{\sc eq}}(k_0)$, with its internal
energy $U_{\text{\sc eff}}^{\star}(k_0) := \<\hat{H}_{\text{\sc
eff}}^{\star}(k_0)\>_{{R}_{\text{\sc eq}}(k_0)}$, being identical to
the internal energy $U_s(k_0) := \<\hat{H}_s\>_{{R}_{\text{\sc
eq}}(k_0)}$ of the coupled oscillator $\hat{H}_s$, as well as its
von-Neumann entropy $S_{\text{\sc eff}}^{\star}(k_0) =
-k_B\,\mbox{Tr}\{\hat{R}_{\text{\sc eq}}(k_0)\,\ln
\hat{R}_{\text{\sc eq}}(k_0)\} = k_B \{v(k_0) + 1/2\}\,\ln \{v(k_0)
+ 1/2\} - k_B \{v(k_0) - 1/2\}\,\ln \{v(k_0) - 1/2\}$ in terms of
$v(k_0) =
\sqrt{\<\hat{q}^2\>_{\beta}(k_0)\cdot\<\hat{p}^2\>_{\beta}(k_0)}/\hbar$,
being identical to that of the coupled oscillator. Here the mass of
the effective oscillator is given by $M_{\text{\sc
eff}}^{\star}(k_0) = \<\hat{p}^2\>_{\beta}(k_0)/U_s(k_0)$, and the
effective spring constant $k_{\text{\sc eff}}^{\star}(k_0) = (k_0 +
\<\hat{p}^2\>_{\beta}(k_0)/\{M\,\<\hat{q}^2\>_{\beta}(k_0)\})/2$.
Subsequently the effective frequency easily follows as
\begin{equation}\label{eq:time-dpt-frequency}
    \omega_{\text{\sc eff}}^{\star}(k_0)\, =\, \frac{1}{2 M}
    \sqrt{\frac{\<\hat{p}^2\>_{\beta}(k_0)}{\<\hat{q}^2\>_{\beta}(k_0)}}\,+\,\frac{k_0}{2}
    \sqrt{\frac{\<\hat{q}^2\>_{\beta}(k_0)}{\<\hat{p}^2\>_{\beta}(k_0)}}\,,
\end{equation}
which also allows us to have
\begin{equation}
    U_{\text{\sc eff}}^{\star}(k_0)\; =\; \omega_{\text{\sc
    eff}}^{\star}(k_0)\,\sqrt{\<\hat{q}^2\>_{\beta}(k_0)\cdot\<\hat{p}^2\>_{\beta}(k_0)}\,.
\end{equation}
Therefore, for the single state $\hat{R}_{\text{\sc eq}}(k_0)$ we
now have two different pictures of the Hamiltonian in consideration,
namely, the coupled oscillator $\hat{H}_s(k_0)$ and its uncoupled
effective counterpart $\hat{H}_{\text{\sc eff}}^{\star}(k_0)$.

Then it can be shown that the effective picture $\hat{H}_{\text{\sc
eff}}^{\star}(k_0)$ is, remarkably enough, exactly in the canonical
thermal equilibrium state $\hat{R}_{\text{\sc eq}}(k_0) \propto
\exp\{-\beta_{\text{\sc eff}}^{\star}(k_0)\cdot\hat{H}_{\text{\sc
eff}}^{\star}(k_0)\}$, where $\beta_{\text{\sc eff}}^{\star}(k_0) =
1/\{k_B\,T_{\text{\sc eff}}^{\star}(k_0)\}$ with the well-defined
effective temperature $T_{\text{\sc eff}}^{\star}(k_0) = \hbar
\omega_{\text{\sc eff}}^{\star}(k_0)/(k_B\,\ln
\{1/\xi_{\beta}(k_0)\})$. Here $\xi_{\beta}(k_0) = \{v(k_0) -
1/2\}/\{v(k_0) + 1/2\}$. From this, it also follows that
\begin{subequations}
\begin{eqnarray}
    \<\hat{q}^2\>_{\beta}(k_0) &=& \frac{\hbar}{2 M_{\text{\sc eff}}^{\star}(k_0)\cdot\omega_{\text{\sc
    eff}}^{\star}(k_0)}\; \coth\left\{\frac{\beta_{\text{\sc eff}}^{\star}(k_0)\cdot\hbar\, \omega_{\text{\sc
    eff}}^{\star}(k_0)}{2}\right\}\label{eq:time-dpt-q-sqr1}\\
    \<\hat{p}^2\>_{\beta}(k_0) &=& \frac{M_{\text{\sc eff}}^{\star}(k_0)\cdot\hbar\, \omega_{\text{\sc
    eff}}^{\star}(k_0)}{2}\; \coth\left\{\frac{\beta_{\text{\sc eff}}^{\star}(k_0)\cdot\hbar\, \omega_{\text{\sc
    eff}}^{\star}(k_0)}{2}\right\}\,.\label{eq:time-dpt-q-sqr2}
\end{eqnarray}
\end{subequations}
As a result, for the quasi-static process
(\ref{eq:density_operator10}) we can take all expressions from Eq.
(\ref{eq:hamiltonian_reversible1}) to (\ref{eq:time-dpt-q-sqr2})
simply with replacement of $k_0 \to k(t)$; e.g., the internal energy
$U_{\text{\sc eff}}^{\star}\{k(t)\} = U_s\{k(t)\} =
\<\hat{p}^2\>_{\beta}\{k(t)\}/2M +
k(t)\,\<\hat{q}^2\>_{\beta}\{k(t)\}/2$, which is surely different
from its non-equilibrium counterpart
${}_{1}\hspace*{-.01cm}{\mathcal U}_s(t)$ in
(\ref{eq:hamiltonian_internal_energy1}) (note that the
time-dependency of the quasi-static quantities comes entirely
through the $k$-value specified by time $t$). Needless to say, in
case that the coupling constants $c_j \to 0$, then
$\hat{H}_{\text{\sc eff}}^{\star}(k) \to \hat{H}_s(k)$ as well as
$T_{\text{\sc eff}}^{\star}(k) \to T$. Also, for the upcoming
numerical analysis it is useful to point out that in the Drude
damping model we substitute $\omega_0^2 \to \omega^2(k) = k(t)/M$
into Eq. (\ref{eq:parameter-change1}), which will give the
expression of the parameter $\Omega(k)$ in terms of $\{\omega(k),
\omega_d, \gamma_o\}$, and then those of $\gamma(k) = \omega_d -
\Omega(k)$ and ${\bf w}_0^2(k) = \{k(t)/M\} \{\omega_d/\Omega(k)\}$,
respectively. And $z_1(k) = \gamma(k)/2 + i {\bf w}_1(k)$ and
$z_2(k) = \gamma(k)/2 - i {\bf w}_1(k)$, where ${\bf w}_1(k) =
\sqrt{\{{\bf w}_0(k)\}^2 - \{\gamma(k)/2\}^2}$.

It is also interesting to consider a temporal behavior of a distance
between the non-equilibrium state $\hat{{\mathcal R}}^{(w)}_1(t)$
and its quasi-static counterpart $\hat{R}_{\text{\sc eq}}\{k(t)\}$.
To do so, we adopt a well-defined measure $D_a^2(t) :=
\text{Tr}\,(\hat{{\mathcal R}}^{(w)}_1(t) - \hat{R}_{\text{\sc
eq}}\{k(t)\})^2$ \cite{GRA98}, which is, independent of the
dimension of the Liouville space, between 0 and 2. With the aid of
\cite{ABS74}
\begin{equation}\label{eq:integral_identity2}
    \textstyle\int_{-\infty}^{\infty} dq\,dq'
    \exp\left\{-a (q + q')^2 - b (q - q')^2 + i c (q^2 - q'^2)\right\}\, =\, \frac{\pi}{\sqrt{4 a b + c^2}}
\end{equation}
we can obtain
\begin{eqnarray}\label{eq:bures1}
    \hspace*{-.5cm}&&D_a^2(t)\, =\, \frac{\hbar}{2}\left(\frac{1}{\sqrt{\<\hat{q}^2\>_{\beta}\,\<\hat{p}^2\>_{\beta}}} + \frac{1}{\sqrt{\<\hat{q}^2\>_{\beta}\{k(t)\}\,\cdot\<\hat{p}^2\>_{\beta}\{k(t)\}}}\right)\, -\\
    \hspace*{-.5cm}&&\frac{2 \hbar \sqrt{B_1(t)\,\<\hat{q}^2\>_{\beta}}}{\sqrt{\left(B_1(t)\,\<\hat{q}^2\>_{\beta}\{k(t)\} + \<\hat{q}^2\>_{\beta}\right) \left(B_1(t)\,\<\hat{p}^2\>_{\beta} +
    \<\hat{p}^2\>_{\beta}\{k(t)\}\right)\,\<\hat{q}^2\>_{\beta} + 4
    \hbar^2\,\Phi_1^2(t)\,\<\hat{q}^2\>_{\beta}\{k(t)\}}}\,.\n
\end{eqnarray}
In Fig. \ref{fig:fig3} this measure for $k(t) = k_0\,(2-e^{-t})$ is
demonstrated for different parameters. Similarly we can also have
\begin{eqnarray}\label{eq:bures2}
    D_{b}^2(t) &:=& \mbox{Tr}\,\{\hat{{\mathcal R}}^{(w)}_1(t) - \hat{{\mathcal R}}(0)\}^2\, =\, \frac{\hbar}{\sqrt{\<\hat{q}^2\>_{\beta}\,\<\hat{p}^2\>_{\beta}}}
    - \frac{2 \hbar \sqrt{B_1(t)}}{\sqrt{\{B_1(t) + 1\}^2\,\<\hat{q}^2\>_{\beta}\,\<\hat{p}^2\>_{\beta} + 4 \hbar^2\,\Phi_1^2(t)}}\n\\
    D_{c}^2(t) &:=& \mbox{Tr}\,\{\hat{R}_{\text{\sc eq}}\{k(t)\} - \hat{R}_{\text{\sc eq}}(k_0)\}^2\, =\, \frac{\hbar}{2}\left(\frac{1}{\sqrt{\<\hat{q}^2\>_{\beta}\,\<\hat{p}^2\>_{\beta}}} +
    \frac{1}{\sqrt{\<\hat{q}^2\>_{\beta}\{k(t)\}\cdot\<\hat{p}^2\>_{\beta}\{k(t)\}}}\right)\n\\
    && - \frac{2 \hbar}{\sqrt{\left(\<\hat{q}^2\>_{\beta} + \<\hat{q}^2\>_{\beta}\{k(t)\}\right) \left(\<\hat{p}^2\>_{\beta} +
    \<\hat{p}^2\>_{\beta}\{k(t)\}\right)}}
\end{eqnarray}
(cf. Fig. \ref{fig:fig4}). Finally it should be stated that all
results in Sect. \ref{sec:instantaneous-eigenvalue} also hold for
the density operator $\hat{{\mathcal R}}^{(w)}_2(t)$ for the mass
variation, simply by replacement of the subscripts $1 \to 2$ and
$k(t) \to M(t)$ of all pertinent parameters.

%%%%%%%%%%%%%%%%%%%%%%%%%%%%%%%%%%%%%%%%%%%%%%%%%%%%%%%%%%%%%%%%%%%%%%%%%%%%%
\section{The second law of thermodynamics}\label{sec:thermodynamics}
%%%%%%%%%%%%%%%%%%%%%%%%%%%%%%%%%%%%%%%%%%%%%%%%%%%%%%%%%%%%%%%%%%%%%%%%%%%%%
%
Based on the results found in the previous sections, we will
explicitly discuss the second law of thermodynamics in the quantum
Brownian oscillator. To address this issue, we need first of all the
first law of thermodynamics
\begin{equation}\label{eq:internal_energy_heat_work1}
    d U_s\, =\, \sum_n\textstyle
    \left(E_n\,dp_n\,+\,p_n\,dE_n\right)\,,
\end{equation}
where $\sum p_n dE_n = \mbox{Tr} (\hat{\rho}\,d\hat{{\mathcal H}}_s)
= \delta{\mathcal W}_s$ corresponds to an amount of work on the
coupled oscillator, and $\sum E_n dp_n = \mbox{Tr} (\hat{{\mathcal
H}}_s\,d\hat{\rho}) = \delta{\mathcal Q}_s$ an amount of heat added
to the oscillator \cite{ALL00}. Next we consider a specific
non-equilibrium process (I), leading to a finite (and so
experimentally measurable), rather than infinitesimal, change in
those thermodynamic quantities, in which the system begins and ends
in thermal equilibrium states but is driven away from thermal
equilibrium at intermediate times. Then an amount of the work along
the process starting with the initial state
(\ref{eq:density_operator1}) is given, with no harm, by
\begin{equation}\label{eq:2nd-law-time-dpt-work0}
    {}_{1}\hspace*{-.05cm}{\mathcal W}_s(t)\, =\, \int_0^{t}
    d\tau\,\dot{k}\,\left\<\frac{\partial {}_{1}\hspace*{-.05cm}\hat{{\mathcal H}}_s(\tau)}{\partial
    k}\right\>_{{\mathcal R}^{(w)}_1(\tau)}\, =\, \frac{1}{2} \int_0^{t}
    d\tau\,\dot{k}\,\<\hat{q}^2\>_{\rho_1(\tau)}
\end{equation}
[cf. Eq. (\ref{eq:expectation_values2})]. Note here that at the end
point $\tau = t$, the system ${}_{1}\hspace*{-.05cm}\hat{{\mathcal
H}}_s(t)$ may not necessarily be in an equilibrium state but relax
to the end equilibrium state $\hat{R}_{\text{\sc eq}}\{k(t)\}$ in
(\ref{eq:density_operator10}). However, no work is performed during
this final stage of thermal relaxation. Then the second law in its
Kelvin-Planck form \cite{CAL85} that this work cannot be less than
its quasi-static counterpart is expressed as
\begin{equation}\label{eq:2nd-law-time-dpt-work1}
    {}_{1}\hspace*{-.05cm}{\mathcal W}_s(t)\, \geq\, W_s\{k(t)\}\,,
\end{equation}
where the work along the quasi-static process
\begin{equation}\label{eq:2nd-law-time-dpt-work11}
    W_s\{k(t)\}\, =\, \int_0^{t}
    d\tau\,\dot{k}\,\left\<\frac{\partial {}_{1}\hspace*{-.05cm}\hat{{\mathcal H}}_s(\tau)}{\partial
    k}\right\>_{R_{{\text{\sc eq}}}\{k(\tau)\}}\, =\, \frac{1}{2} \int_0^t
    d\tau\,\dot{k}\,\<\hat{q}^2\>_{\beta}\{k(\tau)\}\,,
\end{equation}
Fig. \ref{fig:fig5} demonstrates the validity of this inequality and
so that of the second law. Notably, however, based on the fact that
the equilibrium density operator $\hat{R}_{\text{\sc
eq}}\{k(\tau)\}$ is in general not in form of a canonical thermal
state for the coupled oscillator
${}_{1}\hspace*{-.05cm}\hat{{\mathcal H}}_s(\tau)$ in consideration
(rather than its uncoupled counterpart $\hat{H}_{\text{\sc
eff}}^{\star}\{k(\tau)\}$), it can easily be shown that the
quasi-static work $W_s\{k(t)\}$ cannot be interpreted as a
well-defined free energy change of the coupled oscillator
${}_{1}\hspace*{-.05cm}\hat{{\mathcal H}}_s(\tau)$ where $0 \leq
\tau \leq t$.

Here it is also worthwhile to shortly point out that there is an
alternative formulation based on the partition function ${\mathcal
Z}_1\{k(\tau)\} = \mbox{Tr}\,e^{-\beta\,\hat{{\mathcal
H}}_1(\tau)}/\mbox{Tr}_b\,e^{-\beta \hat{H}_b}$, where $\beta =
1/(k_B\,T)$ and the total Hamiltonian $\hat{{\mathcal H}}_1(\tau) =
{}_{1}\hspace*{-.05cm}\hat{{\mathcal H}}_s(\tau) + \hat{H}_b +
\hat{H}_{sb}$
\cite{HAE05,KIM07,FOR06,KIM06,FOR85,HOE05,GEL09,HIL11}. This
immediately leads to the well-defined free energy ${\mathcal
F}_1\{k(\tau)\} = -\ln {\mathcal Z}_1\{k(\tau)\}/\beta$. As
discussed in detail in \cite{KIM10} (the last paragraph of Sect. 3
thereof), however, the free energy ${\mathcal F}_1\{k(\tau)\}$,
containing by definition the coupling-induced ($\hat{H}_{sb}$)
contribution, is not valid for the coupled oscillator
${}_{1}\hspace*{-.05cm}\hat{{\mathcal H}}_s(\tau)$ alone.

Next we discuss the second law in terms of heat. To do so, we first
take into account the internal energy $U_s\{k(\tau)\}
\stackrel{!}{=} U_{\text{\sc eff}}^{\star}\{k(\tau)\}$ of the
coupled oscillator ${}_{1}\hspace*{-.05cm}\hat{{\mathcal
H}}_s(\tau)$ as well as its uncoupled counterpart
$\hat{H}_{\text{\sc eff}}^{\star}\{k(\tau)\}$. The first law of
thermodynamics then tells us that the internal energy change along
the quasi-static process is $U_s\{k(\tau)\}|^t_0 = Q_s\{k(t)\} +
W_s\{k(t)\} \stackrel{!}{=} Q_{\text{\sc eff}}^{\star}\{k(t)\} +
W_{\text{\sc eff}}^{\star}\{k(t)\}$, which is tantamount to
${}_{1}\hspace*{-.05cm}{\mathcal Q}_{\text{\sc eff}}(t) +
{}_{1}\hspace*{-.05cm}{\mathcal W}_{\text{\sc eff}}(t)$ along the
corresponding non-equilibrium process (I) above. Here the
non-equilibrium effective work ${}_{1}\hspace*{-.05cm}{\mathcal
W}_{\text{\sc eff}}(t)$ and its quasi-static counterpart
$W_{\text{\sc eff}}^{\star}\{k(t)\}$ can be obtained directly from
Eqs. (\ref{eq:2nd-law-time-dpt-work0}) and
(\ref{eq:2nd-law-time-dpt-work11}), respectively, with replacement
of the coupled oscillator ${}_{1}\hspace*{-.05cm}\hat{{\mathcal
H}}_s(\tau)$ by its counterpart $\hat{H}_{\text{\sc
eff}}^{\star}\{k(\tau)\}$ such that
\begin{eqnarray}\label{eq:2nd-law-time-dpt-work3}
    {}_{1}\hspace*{-.05cm}{\mathcal W}_{\text{\sc eff}}(t) &=& \frac{1}{2} \int_0^t
    d\tau\,\dot{k}(\tau)\,
    \left(\<\hat{p}^2\>_{\rho_1(\tau)}\, \partial_k
    \frac{1}{M_{\text{\sc eff}}^{\star}\{k(\tau)\}}
    + \<\hat{q}^2\>_{\rho_1(\tau)}\, \partial_k\,k_{\text{\sc
    eff}}^{\star}\{k(\tau)\}\right)\\
    W_{\text{\sc eff}}^{\star}\{k(t)\} &=& \frac{1}{2} \int_0^t
    d\tau\,\dot{k}(\tau)\,
    \left(\<\hat{p}^2\>_{\beta}\{k(\tau)\}\cdot\partial_k
    \frac{1}{M_{\text{\sc eff}}^{\star}\{k(\tau)\}}
    + \<\hat{q}^2\>_{\beta}\{k(\tau)\}\cdot\partial_k\,k_{\text{\sc
    eff}}^{\star}\{k(\tau)\}\right)\n
\end{eqnarray}
where
\begin{subequations}
\begin{eqnarray}
    \partial_k \frac{1}{M_{\text{\sc eff}}^{\star}(k)} &=&
    \frac{\<\hat{q}^2\>_{\beta}(k)}{2\,\<\hat{p}^2\>_{\beta}(k)} + \frac{k(t)}{2}\left\{\frac{\partial_k
    \<\hat{q}^2\>_{\beta}(k)}{\<\hat{p}^2\>_{\beta}(k)}
    - \frac{\<\hat{q}^2\>_{\beta}(k)\cdot\partial_k
    \<\hat{p}^2\>_{\beta}(k)}{\<\hat{p}^2\>_{\beta}^2(k)}\right\}\n\\\\
    \partial_k\,k_{\text{\sc eff}}^{\star}(k) &=&
    \frac{1}{2}\left(1 + \frac{1}{M}\left\{\frac{\partial_k
    \<\hat{p}^2\>_{\beta}(k)}{\<\hat{q}^2\>_{\beta}(k)}
    - \frac{\<\hat{p}^2\>_{\beta}(k)\cdot\partial_k
    \<\hat{q}^2\>_{\beta}(k)}{\<\hat{q}^2\>_{\beta}^2(k)}\right\}\right)\label{eq:2nd-law-time-dpt-work4}
\end{eqnarray}
\end{subequations}
[note the discussion just before Eq. (\ref{eq:time-dpt-frequency})
with replacement of $k_0 \to k$]. And the quasi-static effective
heat $Q_{\text{\sc eff}}^{\star}\{k(t)\}$ can be expressed as
$\int_0^t d\tau\,\dot{k}(\tau)\, T_{\text{\sc
eff}}^{\star}\{k(\tau)\}\cdot\partial_k S_N\{k(\tau)\}$ in terms of
the well-defined effective equilibrium temperature. Here the
von-Neumann entropy $S_N(k)$ is identified with the thermal entropy
of the effective oscillator as
\begin{eqnarray}\label{eq:effective_second_law1}
    \frac{\partial}{\partial k} Q_{\text{\sc eff}}^{\star}(k) &=&
    \frac{1}{2 M_{\text{\sc eff}}^{\star}(k)}\,\frac{\partial}{\partial k} \<\hat{p}^2\>_{\beta}(k)\, +\,
    \frac{k_{\text{\sc eff}}^{\star}(k)}{2}\,\frac{\partial}{\partial k} \<\hat{q}^2\>_{\beta}(k)\n\\
    &=& \frac{\hbar \omega_{\text{\sc eff}}^{\star}(k)}{4} \left\{\frac{\partial}{\partial k} \ln \xi_{\beta}(k)\right\}
    \left\{\text{csch} \frac{\ln \xi_{\beta}(k)}{2}\right\}^2\, =\, T_{\text{\sc eff}}^{\star}(k)\,\frac{\partial}{\partial k}
    S_N(k)\n\\
\end{eqnarray}
[cf. Eqs. (\ref{eq:time-dpt-q-sqr1}) and
(\ref{eq:time-dpt-q-sqr2})].

Now let ${}_{1}\hspace*{-.05cm}{\mathcal W}_{\text{\sc s-eff}}(t) :=
{}_{1}\hspace*{-.05cm}{\mathcal W}_s(t) -
{}_{1}\hspace*{-.05cm}{\mathcal W}_{\text{\sc eff}}(t)$, which can
be interpreted as the work needed for ``switch of picture'' from the
uncoupled effective oscillator to its coupled counterpart along the
non-equilibrium process (I), and its quasi-static counterpart
$W_{\text{\sc s-eff}}\{k(t)\} := W_s\{k(t)\} - W_{\text{\sc
eff}}^{\star}\{k(t)\}$. Substituting these two work functions into
Inequality (\ref{eq:2nd-law-time-dpt-work1}) and applying the above
first law, we can immediately derive a generalized Clausius
inequality
\begin{equation}\label{eq:2nd-law-time-dpt-work2}
    {}_{1}\hspace*{-.05cm}{\mathcal Q}_{\text{\sc eff}}(t)\,
    \leq\, Q_{\text{\sc eff}}^{\star}\{k(t)\}\, +\, {}_{1}\hspace*{-.05cm}\Delta_{\text{\sc s-eff}}(t)\,,
\end{equation}
where ${}_{1}\hspace*{-.05cm}\Delta_{\text{\sc s-eff}}(t) :=
{}_{1}\hspace*{-.05cm}{\mathcal W}_{\text{\sc s-eff}}(t) -
W_{\text{\sc s-eff}}\{k(t)\}$ with
${}_{1}\hspace*{-.05cm}\Delta_{\text{\sc s-eff}}(0) = 0$. Therefore,
in the picture of effective oscillator we hold the standard form of
the Clausius inequality in terms of the well-defined (effective)
temperature, but with the additional term
${}_{1}\hspace*{-.05cm}\Delta_{\text{\sc s-eff}}(t)$. This
inequality can be considered as a consistent generalization of the
Clausius equality $\delta {}_{1}\hspace*{-.05cm}{\mathcal
Q}_{\text{\sc eff}} = T_{\text{\sc eff}}^{\star}\, \partial_k S_N$
valid for the quasi-static process, introduced in \cite{KIM10}.
Obviously, the extra term ${}_{1}\hspace*{-.05cm}\Delta_{\text{\sc
s-eff}}(t)$ identically vanishes in this case. And in the vanishing
coupling limit ($c_j \to 0$), where $T_{\text{\sc eff}}^{\star}(k)
\to T$ as well as both ${}_{1}\hspace*{-.05cm}{\mathcal
W}_{\text{\sc s-eff}}(t) \to 0$ and $W_{\text{\sc s-eff}}\{k(t)\}
\to 0$ leading to ${}_{1}\hspace*{-.05cm}\Delta_{\text{\sc
s-eff}}(t) \to 0$, we can easily recover the ordinary form of the
Clausius inequality in terms of the equilibrium temperature of the
total system. In fact, in the high-temperature limit, where the
thermal fluctuation in the coupled oscillator is predominant to the
system-bath coupling $\hat{H}_{sb}$, the ordinary Clausius
inequality follows as expected. In the low-temperature limit, on the
other hand, the additional term
${}_{1}\hspace*{-.05cm}\Delta_{\text{\sc s-eff}}(t)$ may not be
neglected \cite{COM11}; in Fig. \ref{fig:fig6} we compare the
incomplete Clausius inequality ${}_{1}\hspace*{-.05cm}{\mathcal
Q}_{\text{\sc eff}}(t) \leq Q_{\text{\sc eff}}^{\star}\{k(t)\}$
(with no violation) with its complete counterpart in
(\ref{eq:2nd-law-time-dpt-work2}). As a result, we see that
Inequality (\ref{eq:2nd-law-time-dpt-work2}) is a generalized
Clausius inequality representing the second law in the quantum
Brownian oscillator, without any violation. Finally it should again
be stated that all results in Sect. \ref{sec:thermodynamics} also
hold for the density operator $\hat{{\mathcal R}}^{(w)}_2(t)$,
simply by replacement of the subscripts $1 \to 2$ and $k(t) \to
M(t)$ of all pertinent parameters.

%%%%%%%%%%%%%%%%%%%%%%%%%%%%%%%%%%%%%%%%%%%%%%%%%%%%%%%%%%%%%%%%%%%%%%%%%%%%%
\section{Conclusion}\label{sec:conclusion}
%%%%%%%%%%%%%%%%%%%%%%%%%%%%%%%%%%%%%%%%%%%%%%%%%%%%%%%%%%%%%%%%%%%%%%%%%%%%%
In summary, we have analytically studied non-equilibrium dynamics in
the quantum Brownian oscillator and then systematically discussed
the second law of thermodynamics. We have first derived a closed
expression for the time-dependent reduced density operator of the
coupled oscillator in the weak-coupling limit along the
non-equilibrium process. Based on this density operator, we have
found a generalized Clausius {\em in}equality in terms of the
``effective'' parameters, which indisputably reveals the robustness
of the second law in the quantum regime. In introducing the
effective picture, we reasonably required all thermodynamic
variables to exist and to obey the basic relationships, especially
the first and the second laws. This method, as given, works for the
harmonic oscillator but cannot easily generalized to apply to a
broader class of quantum systems. Therefore the question about the
(rigorous) validity of the second law for such systems remains open.

Our finding can be considered as a consistent generalization of the
Clausius equality valid for the quasi-static process, introduced in
\cite{KIM10}. We believe that this inequality will provide a useful
starting point for later useful discussions of quantum
thermodynamics and quantum information theory within the quantum
Browian oscillator as a prototype of quantum dissipative systems; as
an example, a consistent quantum generalization of the Landauer
principle representing the computational irreversibility may be in
immediate consideration, which can be understood as a simple logical
consequence of the Clausius inequality \cite{LAN61,BEN03}. Lastly,
it is also desirable to numerically study the non-equilibrium
dynamics in this system in the genuine strong-coupling limit as the
next task.

%%%%%%%%%%%%%%%%%%%%%%%%%%%%%%%%%%%%%%%%%%%%%%%%%%%%%%%%%%%%%%%%%%%
% Acknowledgments
%%%%%%%%%%%%%%%%%%%%%%%%%%%%%%%%%%%%%%%%%%%%%%%%%%%%%%%%%%%%%%%%%%%
\section*{Acknowledgments}
The author thanks G. Mahler and A.E. Allahverdyan for useful
discussions. He acknowledges financial support from the Thurgood
Marshall Foundation. He also appreciates all comments and
constructive questions of the referee.

%%%%%%%%%%%%%%%%%%%%%%%%%%%%%%%%%%%%%%%%%%%%%%%%%%%%%%%%%%%%%%%%%%%%%%%%%%%%%
\appendix*\section{: Derivation of the density matrix in Eq. (\ref{eq:density_matrix1})}\label{sec:appendix1}
%%%%%%%%%%%%%%%%%%%%%%%%%%%%%%%%%%%%%%%%%%%%%%%%%%%%%%%%%%%%%%%%%%%%%%%%%%%%%
From comparison of Eqs. (\ref{eq:density_op75}) and
(\ref{eq:density_op8}), it easily follows that $\hat{M} \to
y\partial_y +
\partial_z z$ and $\hat{A} \to yz$ and $\hat{B} \to
\partial_y \partial_z$, and so $\alpha \to 2$ and $m \to
-1$ and $r \to 0$. Let $\lambda \to \tilde{b}_1(t)$ and $\mu \to
\tilde{a}_1(t)/i \hbar\,\tilde{b}_1(t)$ and $\nu \to i
\hbar\,\tilde{c}_1(t)/\tilde{b}_1(t)$. Then we have $d \to 0$ and
$D^2 \to \tilde{a}_1(t)\,\tilde{c}_1(t)/\tilde{b}_1^2(t) - 1$ and
\begin{eqnarray}
    &\displaystyle f \to \frac{1}{i\hbar}\,\frac{\tilde{a}_1(t)}{\tilde{b}_1(t)}\,\frac{\tan\{\tilde{b}_1(t)\,D\}}{D - \tan\{\tilde{b}_1(t)\,D\}}\;\;\; ;\;\;\;
    g \to i\hbar\,\frac{\tilde{c}_1(t)}{\tilde{b}_1(t)}\,\frac{\tan\{\tilde{b}_1(t)\,D\}}{D - \tan\{\tilde{b}_1(t)\,D\}}&\label{eq:density_op16_1}\\
    &\displaystyle \kappa \to \ln\left|\frac{D}{D \cos\{\tilde{b}_1(t)\,D\} - \sin\{\tilde{b}_1(t)\,D\}}\right|\,.&\label{eq:density_op16}
\end{eqnarray}
With the aid of (\ref{eq:density_op8}) and (\ref{eq:commutators_1}),
Eq. (\ref{eq:density_op75}) can then be rewritten as
\begin{equation}\label{eq:density_op9}
    T\,\exp\left\{-\frac{i}{\hbar}\,\hat{J}_1(t)\right\}\, =\,
    \exp\left(\kappa\,y \partial_y\right)\, \exp\left(\kappa\,\partial_z z\right)\, \exp\left(f\,e^{-2\kappa} yz\right)\, \exp\left(g\,\partial_y \partial_z\right)\,.
\end{equation}
Substituting (\ref{eq:density_op9}) into (\ref{eq:density_fkt16})
and then using the identity $e^{c\,x \partial_x} F(x) = F(e^c\,x)$
for a function $F$ \cite{DON07} and $\partial_z z = 1 + z
\partial_z$, we can easily obtain
\begin{equation}\label{eq:density_op10}
    \<q|\hat{{\mathcal R}}^{(w)}_1(t)|q'\>\, =\,
    \left(e^{\kappa}\,e^{f y z}\right)\cdot\left\{e^{g \partial_y \partial_z}\,\<q|\hat{{\mathcal R}}(0)|q'\>\right\}^{y \to e^{\kappa} y}_{z \to e^{\kappa} z}
\end{equation}
where the initial equilibrium density operator in
(\ref{eq:density_operator1}) is then expressed in terms of $y$ and
$z$ as
\begin{equation}\label{eq:density_op11}
    \<q|\hat{{\mathcal R}}(0)|q'\>\, =\, \frac{1}{\sqrt{2 \pi\,\<\hat{q}^2\>_{\beta}}}\,
    \exp\left(-\frac{1}{8\,\<\hat{q}^2\>_{\beta}} y^2 -
    \frac{\<\hat{p}^2\>_{\beta}}{2 \hbar^2} z^2\right)\,.
\end{equation}

Let $e^{g \partial_y
\partial_z}\,\<q|\hat{{\mathcal R}}(0)|q'\> = e^{u
\partial_{\tilde{y}} \partial_{\tilde{z}}}\,\exp\left(-\tilde{y}^2
-\tilde{z}^2\right) =: G(\tilde{y},\tilde{z})$ in Eq.
(\ref{eq:density_op10}) with the aid of (\ref{eq:density_op11}),
where $\tilde{y} := y/\sqrt{8\,\<\hat{q}^2\>_{\beta}}$ and
$\tilde{z} := z \sqrt{\<\hat{p}^2\>_{\beta}/2 \hbar^2}$, and $u :=
(g/4 \hbar) \sqrt{\<\hat{p}^2\>_{\beta}/\<\hat{q}^2\>_{\beta}}$. We
subsequently consider the expansion
\begin{eqnarray}
    G(\tilde{y},\tilde{z}) &=& \sum_{n=0}^{\infty} \frac{1}{n!}\,u^n\,\left(\partial_{\tilde{y}}\,\partial_{\tilde{z}}\right)^n\, \exp\left(-\tilde{y}^2 -
    \tilde{z}^2\right)\label{eq:density_op13}\\
    &=& \sum_{n=0}^{\infty}
    \frac{1}{n!}\,u^n\,H_n(\tilde{y})\,H_n(\tilde{z})\, \exp\left(-\tilde{y}^2
    -\tilde{z}^2\right)\,.\label{eq:density_op13_1}
\end{eqnarray}
Here we used an identity of the Hermite polynomial, $H_n(x) =
(-1)^n\,e^{x^2} (d/dx)^n\,e^{-x^2}$ \cite{ABS74}. The Mehler formula
\cite{MEH38,SZE39}
\begin{equation}\label{eq:mehler1}
    \sum_{n=0}^{\infty} \frac{1}{n!}\,u^n\,H_n(x_1)\,H_n(x_2)\; =\;
    \frac{1}{\sqrt{1 - 4 u^2}}\, \exp\left\{\frac{4 u\,x_1 x_2 - 4 \left(x_1^2 + x_2^2\right) u^2}{1 - 4 u^2}\right\}
\end{equation}
then allows us to have
\begin{equation}\label{eq:density_op14}
    G(\tilde{y},\tilde{z})\; =\; \frac{1}{\sqrt{1 - 4 u^2}}\, \exp\left(\frac{4 u\,\tilde{y} \tilde{z} - \tilde{y}^2 -
    \tilde{z}^2}{1 - 4 u^2}\right)\,.
\end{equation}
From Eqs. (\ref{eq:density_op16}), (\ref{eq:density_op10}) and
(\ref{eq:density_op14}) we can finally obtain Eq.
(\ref{eq:density_matrix1}).

%\twocolumn{

%
%
%\newpage
%
\vspace*{2cm}
Fig.~1: (Color online) $B_1(t)$ versus time $t$ for spring constant
$k(t) = k_0\,(2 - e^{-t})$ and dimensionless temperature $k_B
T/\hbar \omega_0$ [cf. Eq. (\ref{eq:density_matrix21})]; dashdot: $T
= 10$ (high temperature), and solid: $T = 0.01$ (low temperature).
Here $\hbar = k_B = \omega_0 = M = k_0 = \gamma_0 = 1$, and
$\omega_d = 5$. Since this spring constant exponentially decays with
time, the coefficients $\tilde{a}_1(t), \tilde{b}_1(t)$ and
$\tilde{c}_1(t)$ in
(\ref{eq:density_op761})-(\ref{eq:density_op763}) should fast
converge, and so the infinite sum over $n$ in
(\ref{eq:exponential_identity1}) can be replaced by a finite sum
with the upper bound $N$ not necessarily large enough. This can be
verified here by considering $N = 1, 2, 3$ and $4$. From top to
bottom (at $t = 10$), (black dashdot: $N = 1$), (blue dashdot: $N =
3$), (red dashdot: $N = 4$), and (green dashdot: $N = 2$); (green
solid: $N = 2$), (red solid: $N = 4$), (blue solid: $N = 3$), and
(black solid: $N = 1$). From this numerical result with an
oscillating and fast converging behavior of $B_1(t)$ with increasing
$N$, we adopt $\{B_1(t)|_{N=3} + B_1(t)|_{N=4}\}/2$ as a numerical
fitting of $B_1(t)$, which will be used for later numerical
studies.\vspace*{.7cm}

\begin{figure}[htb]
\centering\hspace*{-1.5cm}{\includegraphics[scale=0.9]{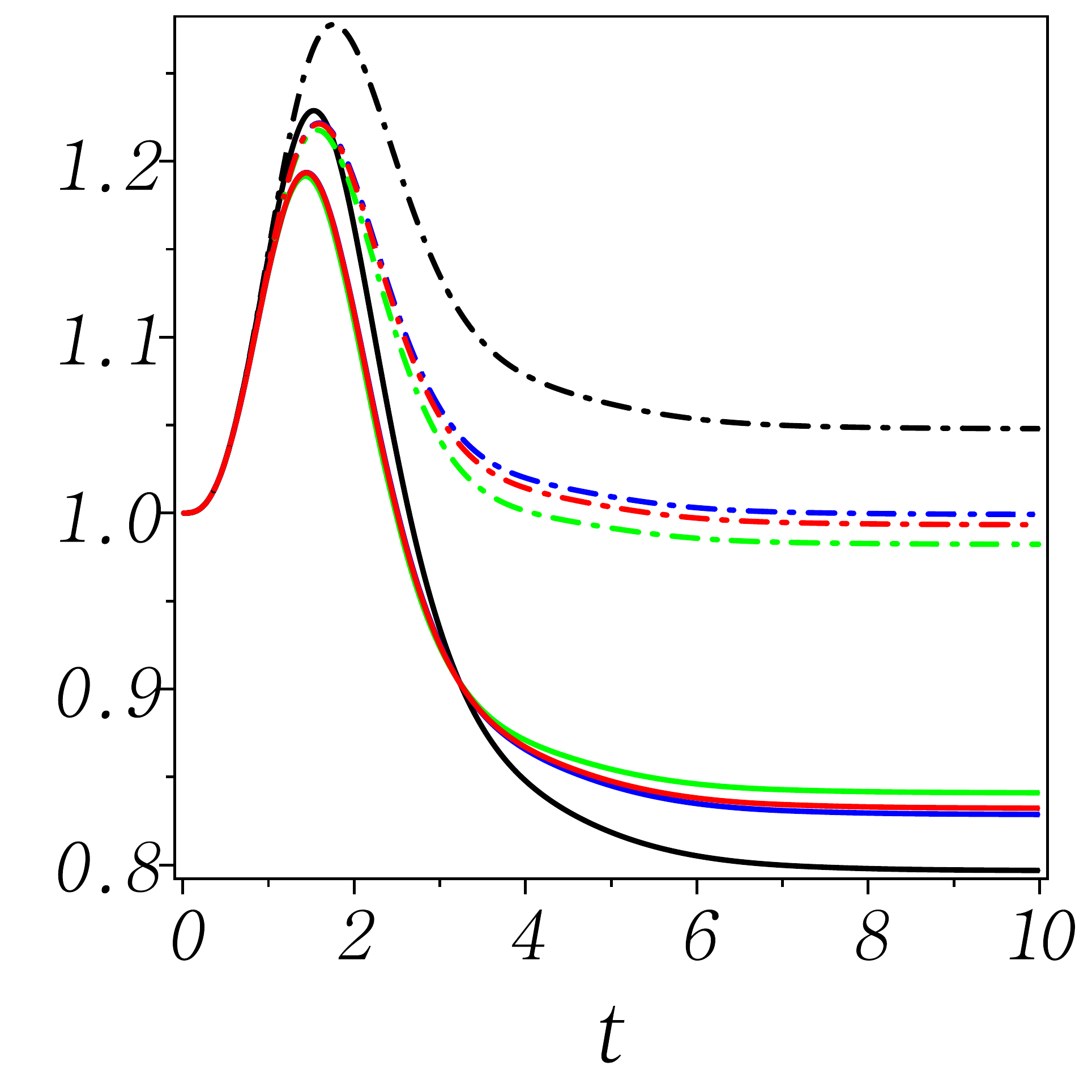}
\caption{\label{fig:fig1}}}
\end{figure}

\begin{figure}[htb]
\centering\hspace*{-1.5cm}{\includegraphics[scale=0.9]{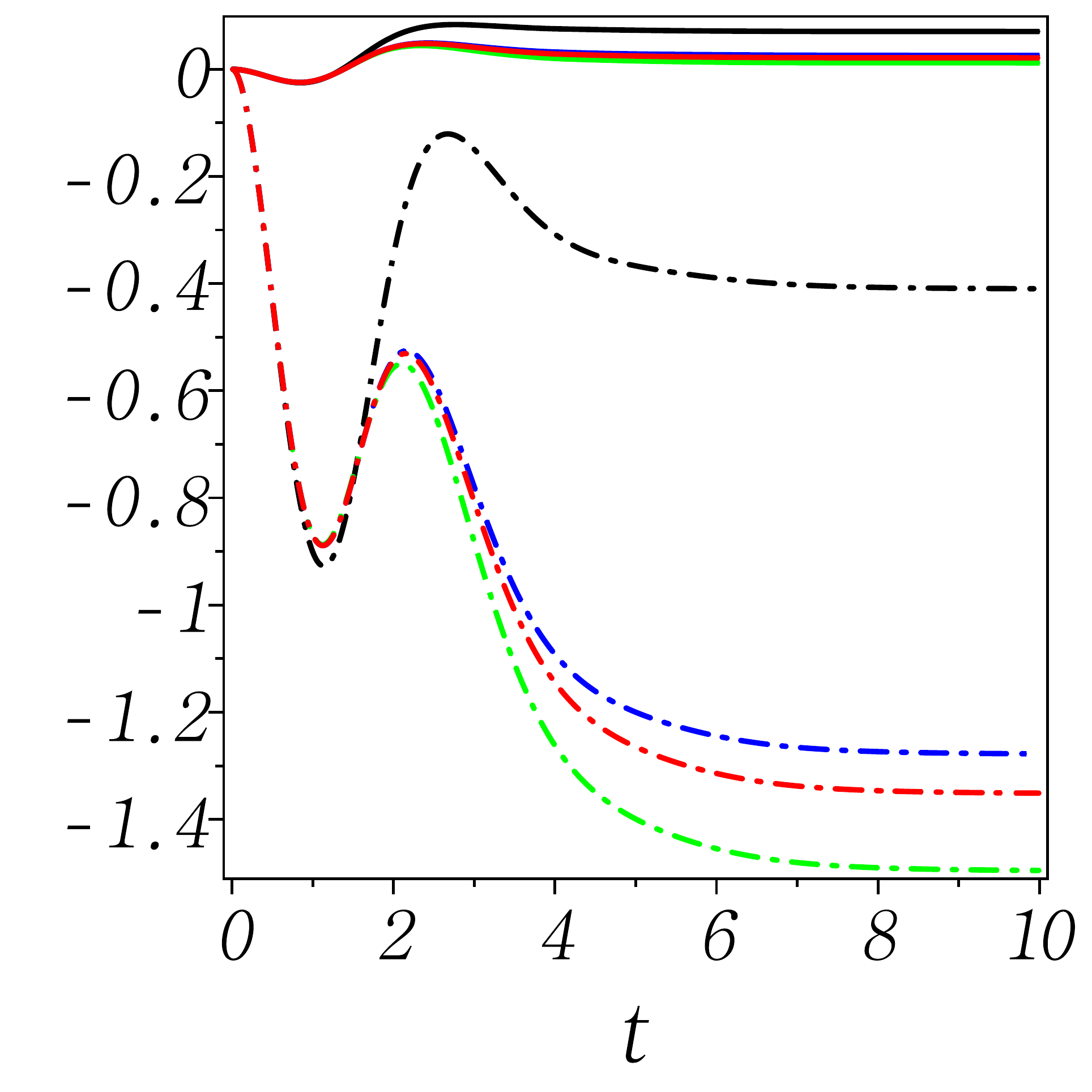}
\caption{\label{fig:fig2}}}
\end{figure}

\begin{figure}[htb]
\centering\hspace*{-1.5cm}{\includegraphics[scale=0.9]{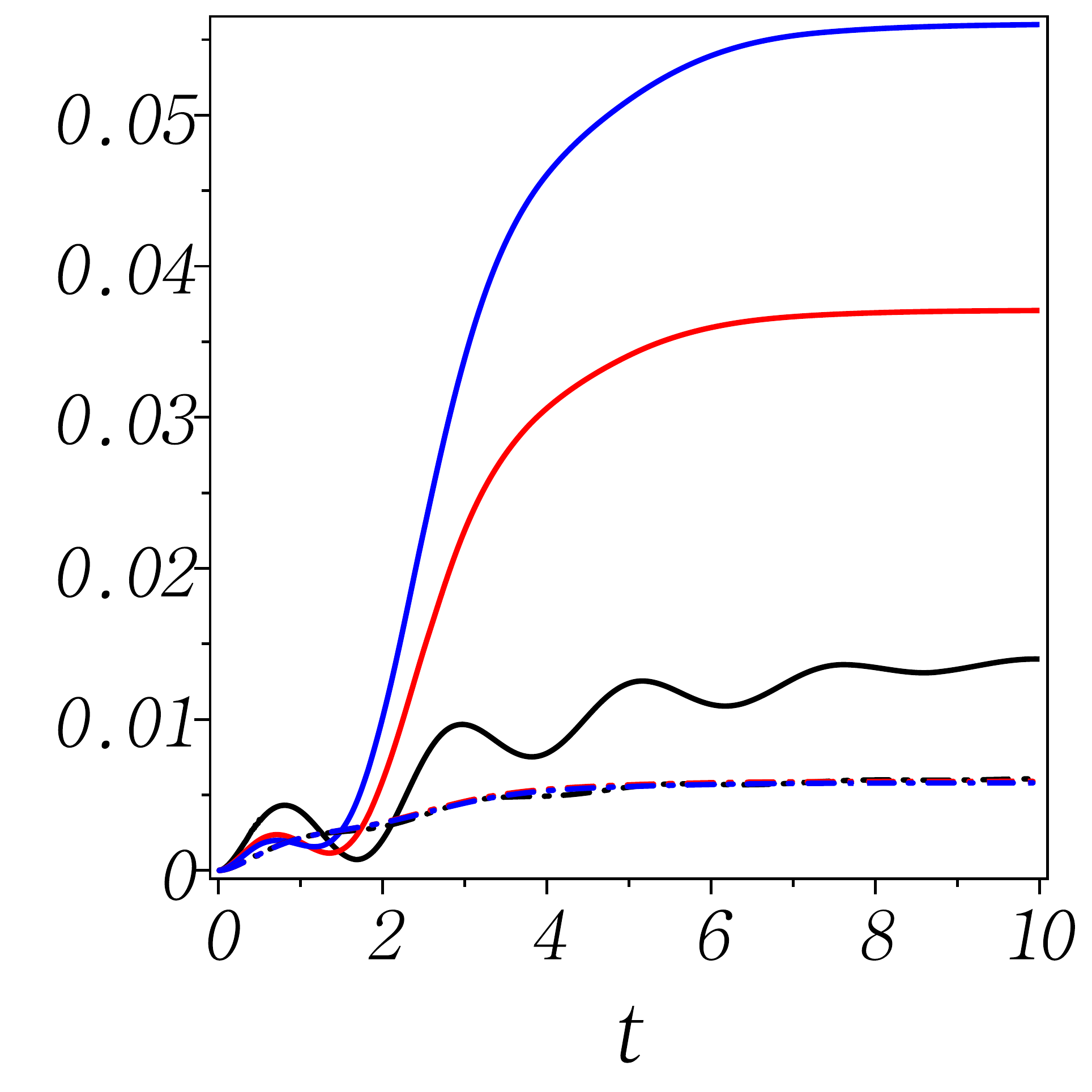}
\caption{\label{fig:fig3}}}
\end{figure}

\begin{figure}[htb]
\centering\hspace*{-1.5cm}{\includegraphics[scale=0.9]{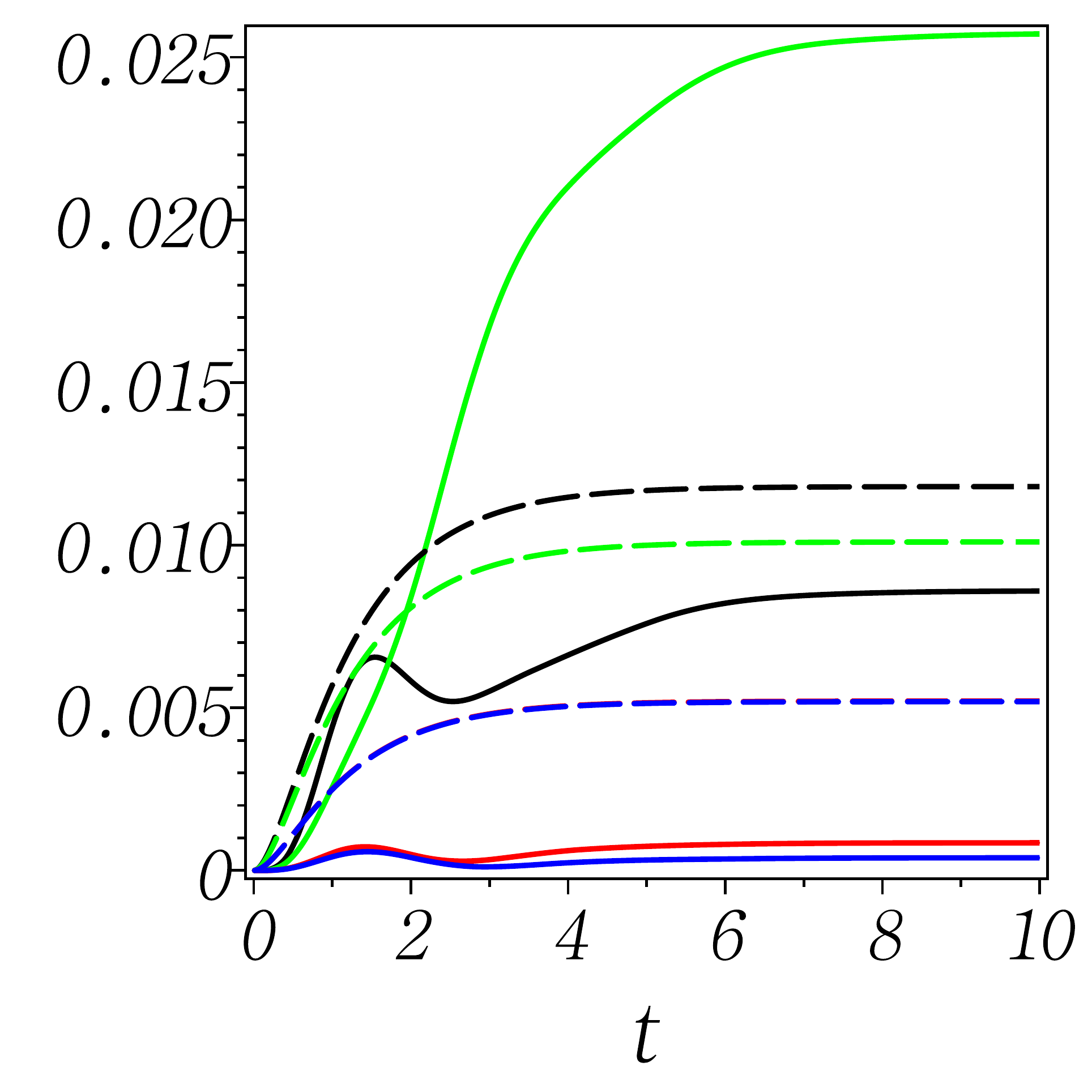}
\caption{\label{fig:fig4}}}
\end{figure}

Fig.~2: (Color online) $\Phi_1(t)$ versus time $t$ for spring
constant $k(t) = k_0\,(2 - e^{-t})$ and dimensionless temperature
$k_B T/\hbar \omega_0$ [cf. Eq. (\ref{eq:density_matrix22})];
dashdot: $T = 10$ (high temperature), and solid: $T = 0.01$ (low
temperature). Here $\hbar = k_B = \omega_0 = M = k_0 = \gamma_0 =
1$, and $\omega_d = 5$. As for Fig. \ref{fig:fig1}, we consider $N =
1, 2, 3$ and $4$. From top to bottom (at $t = 10$), (black solid: $N
= 1$), (blue solid: $N = 3$), (red solid: $N = 4$), and (green
solid: $N = 2$); (black dashdot: $N = 1$), (blue dashdot: $N = 3$),
(red dashdot: $N = 4$), and (green dashdot: $N = 2$). From this
numerical result with an oscillating and fast converging behavior of
$\Phi_1(t)$ with increasing $N$, we adopt $\{\Phi_1(t)|_{N=3} +
\Phi_1(t)|_{N=4}\}/2$ as a numerical fitting of $\Phi_1(t)$, which
will be used for later numerical studies. Especially for $T = 0.01$,
$\Phi_1(t)|_{N=3} \approx \Phi_1(t)|_{N=4}$ already.\vspace*{.7cm}

Fig.~3: (Color online) $D_a^2(t)$ versus time $t$ for spring
constant $k(t) = k_0\,(2 - e^{-t})$ and dimensionless temperature
$k_B T/\hbar \omega_0$ [cf. Eq. (\ref{eq:bures1})]; dashdot: $T =
10$ (high temperature), and solid: $T = 0.01$ (low temperature).
Here $\hbar = k_B = \omega_0 = M = k_0 = \gamma_0 = 1$. From top to
bottom (at $t = 10$), (blue solid: cut-off frequency $\omega_d =
15$), (red solid: $\omega_d = 5$), and (black solid: $\omega_d =
1$); (black dashdot: $\omega_d = 1$), (red dashdot: $\omega_d = 5$),
and (blue dashdot: $\omega_d = 15$). The three curves at $T = 10$
are almost identical.\vspace*{.7cm}

Fig.~4: (Color online) \{$D_b^2(t)$, solid\} and \{$D_c^2(t)$,
dash\} versus time $t$ for spring constant $k(t) = k_0\,(2 -
e^{-t})$ and dimensionless temperature $k_B T/\hbar \omega_0$ [cf.
Eq. (\ref{eq:bures2})]. Here $\hbar = k_B = \omega_0 = M = k_0 =
\gamma_0 = 1$. Solid: from top to bottom (at $t=10$), (green: $T =
0.01$ and $\omega_d = 15$), (black: $T = 0.01$ and $\omega_d = 5$),
(red: $T = 10$ and $\omega_d = 5$), and (blue: $T = 10$ and
$\omega_d = 15$). Dash: from top to bottom (at $t=10$), (black: $T =
0.01$ and $\omega_d = 5$), (green: $T = 0.01$ and $\omega_d = 15$),
and (red: $T = 10$ and $\omega_d = 5$) $\approx$ (blue: $T = 10$ and
$\omega_d = 15$).\vspace*{.7cm}

Fig.~5: (Color online) ${}_{1}\hspace*{-.05cm}{\mathcal W}_s(t) -
W_s\{k(t)\} =: y_5 \geq 0$ versus time $t$ for spring constant $k(t)
= k_0\,(2 - e^{-t})$ and dimensionless temperature $k_B T/\hbar
\omega_0$ [cf. Eq. (\ref{eq:2nd-law-time-dpt-work1})]; dash: $T =
10$ (high temperature), and solid: $T = 0.01$ (low temperature).
Here $\hbar = k_B = \omega_0 = M = k_0 = \gamma_0 = 1$. From top to
bottom (at $t = 10$), (black dash: $\omega_d = 15$), (red dash:
$\omega_d = 5$), and (blue dash: $\omega_d = 1$) for $y_5/20$, and
then (black solid: $\omega_d = 15$), (red solid: $\omega_d = 5$),
and (blue solid: $\omega_d = 1$) for $y_5$.\vspace*{.7cm}

\begin{figure}[htb]
\centering\hspace*{-1.5cm}{\includegraphics[scale=0.9]{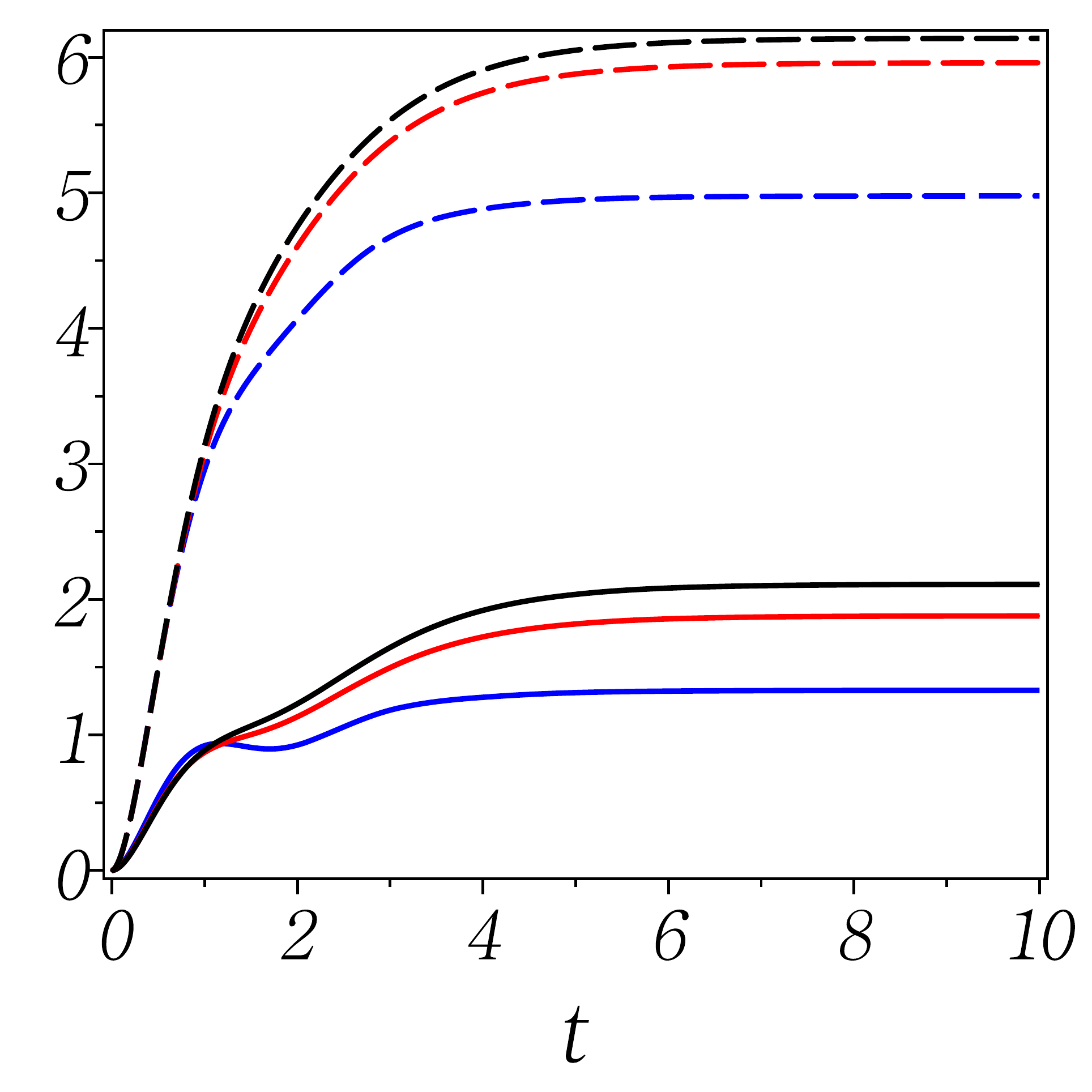}
\caption{\label{fig:fig5}}}
\end{figure}

\begin{figure}[htb]
\centering\hspace*{-1.5cm}{\includegraphics[scale=0.9]{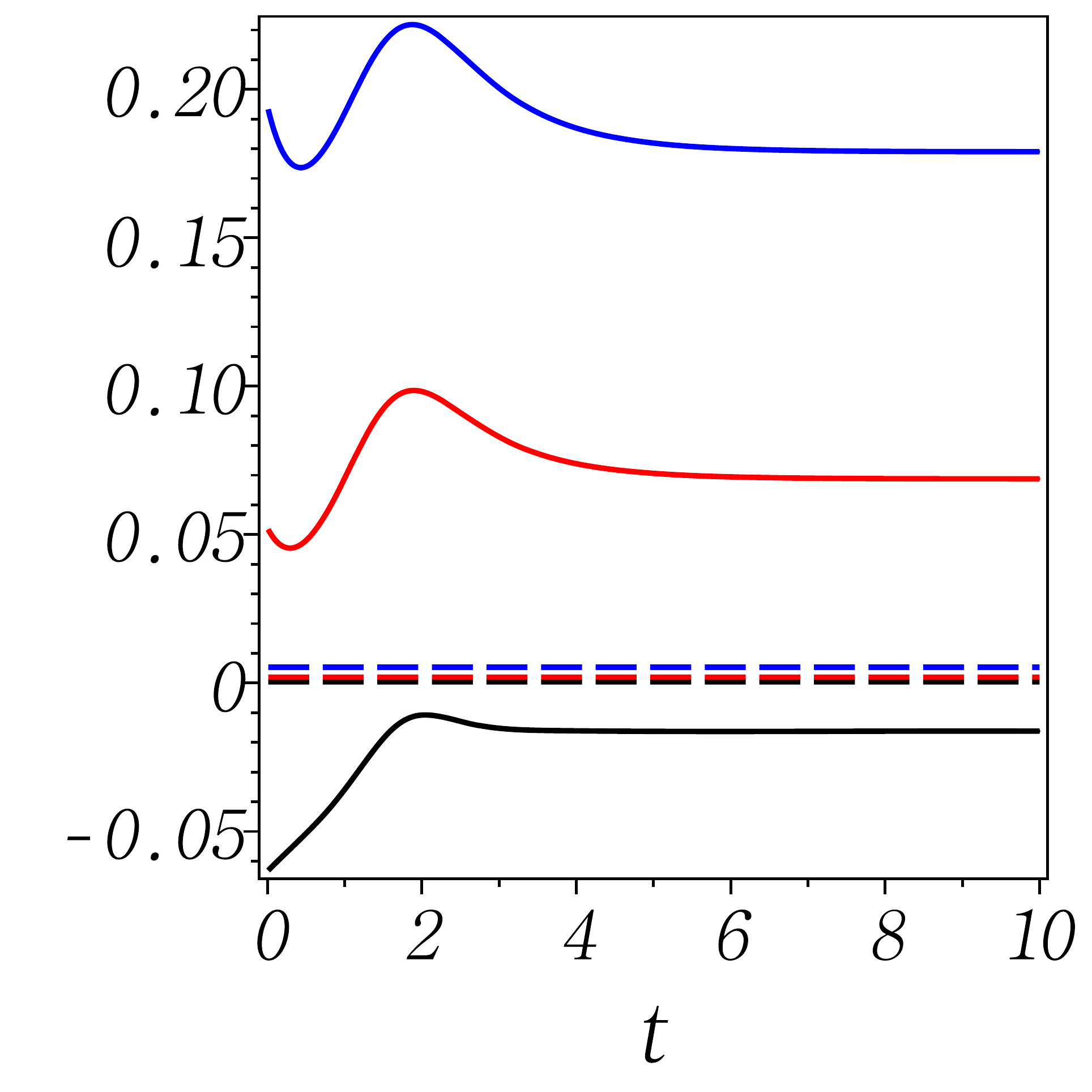}
\caption{\label{fig:fig6}}}
\end{figure}

Fig.~6: (Color online) $\{{\mathcal I}_{\text{\sc eff}}(t) -
{\mathcal J}_{\text{\sc eff}}(t)\}/{\mathcal J}_{\text{\sc eff}}(t)
=: y_6$ versus time $t$ for spring constant $k(t) = k_0\,(2 -
e^{-t})$ and dimensionless temperature $k_B T/\hbar \omega_0$. Here
${\mathcal I}_{\text{\sc eff}}(t) := Q_{\text{\sc
eff}}^{\star}\{k(t)\} - {}_{1}\hspace*{-.05cm}{\mathcal
Q}_{\text{\sc eff}}(t)$, and ${\mathcal J}_{\text{\sc eff}}(t) :=
Q_{\text{\sc eff}}^{\star}\{k(t)\} - {}_{1}\hspace*{-.05cm}{\mathcal
Q}_{\text{\sc eff}}(t) + {}_{1}\hspace*{-.05cm}\Delta_{\text{\sc
s-eff}}(t)$ [cf. Eq. (\ref{eq:2nd-law-time-dpt-work2})], which is
identical to ${}_{1}\hspace*{-.05cm}{\mathcal W}_s(t) - W_s\{k(t)\}$
in Fig. \ref{fig:fig5}; dash: $T = 10$ (high temperature), and
solid: $T = 0.01$ (low temperature). And $\hbar = k_B = \omega_0 = M
= k_0 = \gamma_0 = 1$. Solid: from top to bottom, (blue: $\omega_d =
15$), (red: $\omega_d = 5$), and (black: $\omega_d = 1$). Dash: in
the same order. The three curves at $T = 10$ almost vanish, as
expected. At $T = 0.01$, we see that
${}_{1}\hspace*{-.05cm}\Delta_{\text{\sc s-eff}}(t) < 0$ for
$\omega_d = 5$ and $15$, immediately leading to no violation of the
incomplete inequality ${\mathcal I}_{\text{\sc eff}}(t) \geq 0$; on
the other hand, ${}_{1}\hspace*{-.05cm}\Delta_{\text{\sc s-eff}}(t)
> 0$ for $\omega_d = 1$. However, from the fact that $y_6 \geq -1$ in
this case, there is still no violation of ${\mathcal I}_{\text{\sc
eff}}(t) \geq 0$.\vspace*{.7cm}

\begin{thebibliography}{1}
\bibitem{COM11} In fact, in the strong-coupling limit, in which the effective temperature $T_{\text{\sc
eff}}^{\star}(k)$ heavily differs from the total-system temperature
$T$, we can easily deduce from the discussion for Eqs.
(\ref{eq:2nd-law-time-dpt-work3})-(\ref{eq:2nd-law-time-dpt-work2})
that the Clausius inequality in this limit should also include a
(non-negligible) extra term corresponding to
${}_{1}\hspace*{-.05cm}\Delta_{\text{\sc s-eff}}(t)$ in
(\ref{eq:2nd-law-time-dpt-work2}) derived in the weak-coupling
limit.
\end{thebibliography}
\end{document}